\documentclass[epj]{svjour}
\usepackage{amssymb}
\usepackage{amsbsy}
\usepackage{graphicx}
\usepackage{texdraw}
\begin{document}
\sloppy
\title{Bulk singularities at critical end points: a field-theory analysis}
\author{H.\ W.\ Diehl\thanks{e-mail: phy300{\@}theo-phys.uni-essen.de} and M.\ Smock\thanks{Present address: CH-4641 H{\"a}gendorf, Switzerland}%
}                
\institute{Fachbereich Physik, Universit{\"a}t Essen,
 D-45117 Essen, Federal Republic of Germany
}
\date{Received: / Revised version: March 29, 2001 }
\mail{H.\ W.\ Diehl, Fachbereich Physik, Universit{\"a}t-GH Essen,
D-45117 Essen, Germany}
\abstract{
A class of continuum models with a critical end point is considered
whose Hamiltonian ${\mathcal{H}}[\phi,\psi]$ involves two densities:
a primary order-parameter field, $\phi$, and a secondary
(noncritical) one, $\psi$. Field-theoretic methods (renormalization group
results in conjunction with functional methods) are used to give a
systematic derivation of singularities occurring at
critical end points. Specifically, the thermal
singularity $\sim|{t}|^{2-\alpha}$ of the first-order line on which
the disordered or ordered
phase coexists with the noncritical spectator phase,
and the coexistence singularity $\sim |{t}|^{1-\alpha}$ or
$\sim|{t}|^{\beta}$
of the secondary density $\langle\psi\rangle$ are derived.
It is clarified how the
renormalization group (RG) scenario found in position-space RG
calculations, in which the critical end point and the critical line are
mapped onto two separate fixed points ${\mathcal P}_{\mathrm{CEP}}^*$.
and ${\mathcal P}_{\lambda}^*$ translates into field theory.
The critical RG eigenexponents of ${\mathcal P}_{\mathrm{CEP}}^*$
and ${\mathcal P}_{\lambda}^*$ are shown to
match. ${\mathcal P}_{\mathrm{CEP}}^*$ is demonstrated
to have a  discontinuity eigenperturbation (with
eigenvalue $y=d$), tangent to the unstable trajectory that emanates
from ${\mathcal P}_{\mathrm{CEP}}^*$ and leads to ${\mathcal P}_{\lambda}^*$.
The nature and origin
of this eigenperturbation as well as the role redundant operators
play are elucidated. The results validate that
the critical behavior at the end point is the same as on the critical line.
 % end of PACS codes
} %end of abstract
\maketitle
\section{Introduction}\label{sec:intro}

Critical end points are widespread in nature. They occur when a
line of critical temperatures (or lambda line) $T_{\mathrm{c}}(g)$, depending on
a nonordering field $g$ such as chemical potential
or pressure, terminates at a line $g_\sigma(T)$
of discontinuous phase transitions
\cite{rem1,Gri73,FB91,CL95}.
In the past decades plenty of
experimental and theoretical studies
have been made
in which critical end points were encountered
\cite{Gri73,FB91,CL95,Fis89,vKS79,Wid77,ZAGJ82,Wid85,FU90a,FU90b,%
Law94,Wil97a,Wil97b,%
WSN98,BF91b,SvK70,GD93,RCC92,DL89b,%
Bar91,BW76,Fis91,GS88,KGYF81,MS76,RW85,Sob71,TW83,PPK73}. Yet
they have been rarely investigated for their own sake.
This may be due to the physically appealing and widely
accepted, though seldom carefully checked,
expectation that the critical phenomena at such an end point
should not differ in any significant way from critical phenomena along
the critical line $T_{\mathrm{c}}(g)$ \cite{Gri73}.

\begin{figure}[htb]
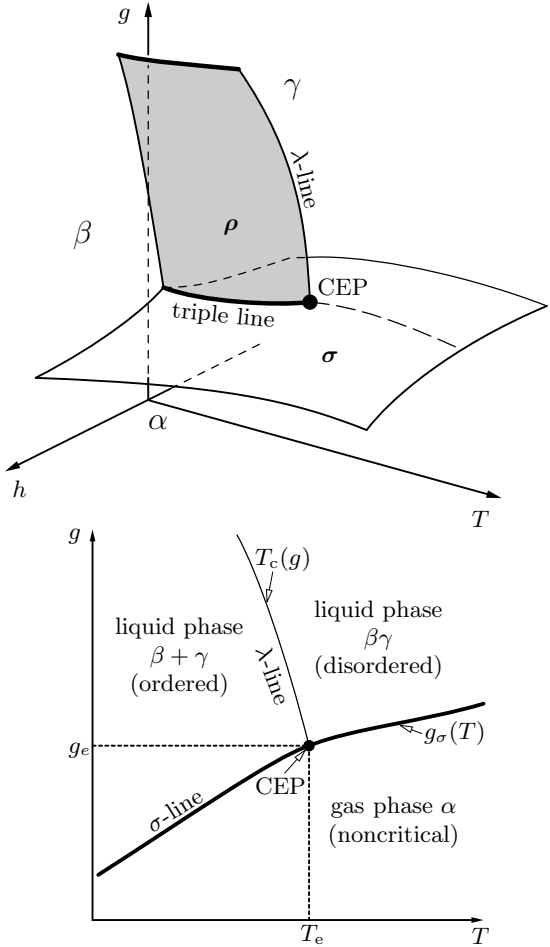

\begin{center}
\begin{texdraw}
\drawdim mm
\move (3 1.5) 
 \arrowheadtype t:F \arrowheadsize l:2 w:1 \ravec(-22 -11)
\move(0 0) \rlvec(0 3) \rmove(0 43) \ravec(0 7)
\move(0 0) \ravec (47 -13)
\move (29 -4) \clvec (19 0)(10 2)(-15 3)
\clvec(-6 7.5)(0 12.5)(2 15)
\move(29 -4) \clvec(32 0)(40 7)(53 12.5)
\move(2 15)
\clvec(0 28)(-2 40)(-4 46) 
\clvec(0 45)(7 44.5)(12 44) 
\clvec(20 32.5)(21 23)(21.5 13)
\clvec(17 12.5)(7 13)(2 15)
\lfill f:0.8 \move(21.5 13)  \linewd 0.6 
\clvec(17 12.5)(7 13)(2 15)
\move(-4 46) \clvec(0 45)(7 44.5)(12 44)
\move (3 1.5)\linewd 0.2 \lpatt (1 1) \rlvec(12 6)
\move(0 1.5) \rlvec(0 43)
\move(2 15)
\clvec(7 15)(10 16)(19 19) \rlvec (2 0)
\lpatt() \clvec(27 19.5)(40 17.5)(53 12.5)
\move(21.5 13) \fcir f:0 r:1 \lpatt(2.5 1) \clvec(25 12.7)(30 12)(41 7)
\htext(-4 50){$g$} \htext(43 -17){$T$}
\htext(-18 -13){$h$}
\rtext td:-76 (19 33){$\lambda$-line}
\htext(23 5){$\boldsymbol{\sigma}$}
\htext(10 22){$\boldsymbol{\rho}$}
\htext(-10 20){\large$\beta$}
\htext(18 40){\large$\gamma$}
\htext(0 -4){\large$\alpha$}
\rtext td:-6 (3 11){triple line}
\htext(22.5 14){CEP} 
\end{texdraw}

\begin{texdraw}
\drawdim mm \setunitscale 0.8 \move (0 0)
\arrowheadtype t:F \arrowheadsize l:2 w:1 \ravec(65 0)
\move (0 0) \ravec (0 65)
\move(0 29) \lpatt (0.5 0.5) \rlvec(36 0) \rlvec(0 -29)
\move(36 29) \lpatt() \linewd 0.25 \clvec (32.5 43)(27.5 57.8)(24 64)
\htext(-4 62.5){$g$} \htext(-4 27){$g_e$} \htext(34.5 -4){$T_{\mathrm{e}}$}
\htext(63 -4){$T$}
\htext(0 37){\small\begin{minipage}[t]{23mm}%
\begin{center}liquid phase\\ $\beta +\gamma$\\(ordered)\end{center}\end{minipage}}
\htext(30 40){\small\begin{minipage}[t]{28mm}\begin{center}liquid phase\\
{$\beta\gamma$}\\(disordered)\end{center}\end{minipage}}
\htext(32.5 12){\small\begin{minipage}[t]{28mm}\begin{center}gas phase
$\alpha$\\(noncritical)\end{center}\end{minipage}}
\move(65 36)\lpatt()\linewd 0.6 \clvec (55 33)(43.5 32)(36 29)
\fcir f:0 r:1
\clvec (30 27)(15 16.5)(1 7.5)
\htext(55 29){\small$g_\sigma(T)$}
\rmove(-0.3 2)\lpatt() \linewd 0.1 
\arrowheadtype t:T \arrowheadsize l:2 w:1 \ravec (-3.6 1.4)
\htext(27 58){\small$T_{\mathrm{c}}(g)$} \rmove (3 -0.1) \ravec (-1 -5)
\htext(27 21){\small CEP} \rmove (5 3.5)\avec (35.3 28.3)
%\rtext td:-62 (20 63){$\lambda$-line}
\rtext td:-70 (27 45){$\lambda$-line}
\rtext td:32 (10 14.5){$\sigma$-line}
\end{texdraw}
\end{center}
\caption{Upper drawing: Schematic phase diagram showing,
in a field space,
portions of the coexistence surfaces $\boldsymbol{\rho}$ and
$\boldsymbol{\sigma}$ for a nonsymmetric binary mixture
with a liquid-liquid critical line $\lambda$ and a  critical end point
$\mathrm{CEP}$ (full circle). 
The lower drawing represents the projection
of the phase diagram on the $gT$-plane.
Along the triple line the ordered phases
$\beta$ (A-rich) and $\gamma$ (A-poor) coexist
with the spectator phase $\alpha$ (vapor).
 }
\label{fig:scpd}
\end{figure}

However, Fisher and collaborators have pointed out recently
\cite{Fis89,FU90a,FU90b,FB91}
that even the {\em bulk thermodynamics\/}
of a critical end point  should display new critical singularities,
not observable on the critical line. For concreteness, let us
consider the simple critical end point situation
depicted in Fig.~\ref{fig:scpd} of a binary
fluid mixture, whose critical line, $\lambda$,%
\footnote{For the sake of simplicity, we shall henceforth refer to the critical line
$\lambda$ briefly as the `$\lambda$-line', a term usually reserved for the critical
line of normalfluid-to-superfluid transitions of He. Since we shall explicitly consider
only the case of a scalar order parameter $\phi$ (even though parts of our analysis
can  be generalized in a straightforward fashion to more general cases with an
multi-component order parameter),
no confusion should arise from this slight abuse of terminology.}
is restricted to values $g\ge g_e$ and ends at
the end point temperature $T_{\mathrm{e}}=T_{\mathrm{c}}(g_e)$.
We assume that the critical heat singularity at constant $g>g_e$
on the critical line can be
written as  $A_\pm(g)\,\left|T-T_{\mathrm{c}}(g)\right|^{-\alpha}$, where $\alpha>0$
in $d<4$ dimensions. Fisher's assertion then is that
the first-order phase boundary should vary as
\begin{equation}\label{singpb}
g_\sigma(T)\approx g^{\mathrm{reg}}_\sigma(T)-\frac{X^0_\pm}{(2-\alpha)(1-\alpha)}
\,\left|{t}\right|^{2-\alpha}\end{equation}
as ${t}\equiv  (T-T_{\mathrm{e}})/ T_{\mathrm{e}}\to \pm 0$. Here $g^{\mathrm{reg}}_\sigma(T)$ is a regular background term, and
$X^0_+/X^0_-$ should be equal to the universal (and hence $g$
independent) ratio $A_+/A_-$.

Equation (\ref{singpb}) has been derived by using general thermodynamic
arguments in conjunction with the phenomenological theory of scaling
\cite{Fis89,FB91,FU90a}. A straightforward extension of such reasoning
reveals that the thermodynamic density conjugate to $g$,
which in the present case
can be identified as the total particle density $\varrho_{\mathrm{tot}}$,
displays nonanalytic behavior of the form
\begin{equation}\label{singdens}
 \varrho^{\mathrm{sing}}_{\mathrm{tot}}\equiv \varrho_{\mathrm{tot}}-
\varrho^{\mathrm{reg}}_{\mathrm{tot}}(T)\approx 
U^0_\pm\,\left|{t}\right|^{\beta}
+V^0_\pm\,\left|{t}\right|^{1-\alpha}
\end{equation}
as the end point is approached
along  the liquid side of the liquid-vapor
coexistence curve from ${t}>0$ or
${t}<0$ \cite{Wil97a,Wil97b}.
Here $\varrho^{\mathrm{reg}}_{\mathrm{tot}}(T)$ is regular at $T_{\mathrm{e}}$. Further,
the amplitude
$U^0_\pm$ vanishes for symmetric binary fluids whose properties are
invariant with regard to simultaneous
interchange of the two constituents A and B
and their respective chemical potentials $\mu_{\mathrm{A}}$ and $\mu_{\mathrm{B}}$.

The  $|{t}|^{2-\alpha}$ singularity of (\ref{singpb}) has been confirmed
for exactly solvable spherical models \cite{BF91b}  and
checked by Monte Carlo
calculations \cite{Wil97a,Wil97b}; the $|{t}|^{1-\alpha}$ singularity
of (\ref{singdens})
is consistent with the jump in the slope of $\varrho_{\mathrm{tot}}(T)$
found in mean field and
density functional calculations
\cite{RCC92,GD98} and
 has also been seen in Monte Carlo simulations
\cite{Wil97a,Wil97b,WSN98}.

Despite these advances the present state of the theory is anything but
satisfactory. As can be seen from the  thorough discussion given in Ref.\ \cite{FB91},
a number of challenging problems exist.
First, a systematic derivation of the above-mentioned
singularities within the framework of the modern
field-theoretic renormalization group (RG)
approach is lacking.
We are aware of only a single field-theoretic analysis of
critical end-point behavior that goes beyond Landau theory:
the $(\epsilon=4-d)$-expansion study of
a scalar  $\phi^8$ model with a negative $\phi^6$ term
by Ziman {\it et al.\/} \cite{ZAGJ82}. In their one-loop calculation they found
that both the critical line and the end point were controlled by the same, standard $O(\epsilon)$ fixed point. However, this model has very special properties:
The first-order phase boundary does {\em not\/} extend into the disordered phase;
as the critical end point is approached from the disordered phase, the
order parameter $\langle\phi\rangle$ becomes critical {\em and\/} exhibits a
jump to a nonvanishing value upon entering the ordered phase;
no critical fluctuations occur in the ordered phase.
Thus the model clearly does not reflect the typical critical
end point situation
in which the two-phase coexistence surface $\boldsymbol{\rho}$
bounded by the critical line $T_{\mathrm{c}}(g)$ meets the
spectator phase boundary $\boldsymbol{\sigma}$ in a triple line
(see Fig.~\ref{fig:scpd}).

\begin{figure}[htb]
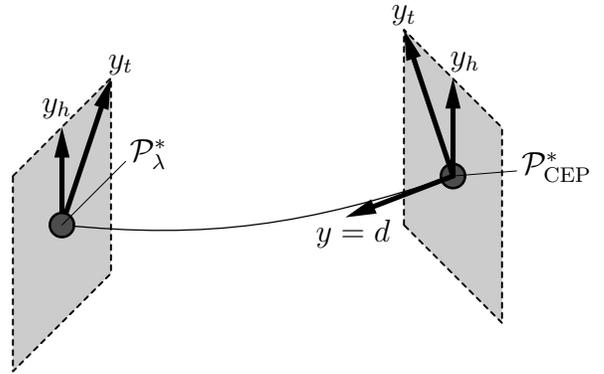

%
%The figure is generated here via texdraw%%%
\begin{center}
\begin{texdraw}
\drawdim mm \setunitscale 0.65
\move (120 0)
\lpatt(1 1)
\rlvec (20 20)
\rlvec (0 -40)
\rlvec (-20 -20)
\rlvec (0 40) \lfill f:0.8
\rmove (80 -10)
\rlvec (0 40)
\rlvec (20 -20)
\rlvec (0 -40)
\rlvec (-20 20)
\lfill f:0.8
\rmove (10 10)
\fcir f:0.3 r:2.5
\lpatt() \linewd 0.5  \lcir r:2.5 
\linewd 1
\arrowheadtype t:F \ravec(0 20)
\rmove (0 -20)
\arrowheadtype t:F \ravec(-10 30)
\rmove(10 -30)
\linewd 0.25
\clvec (180 -10)(160 -13)(130 -10)
\linewd 1
\arrowheadtype t:F \ravec(0 20)
\rmove (0 -20)
\arrowheadtype t:F \ravec(10 30)
\rmove(-10 -30)
\fcir f:0.3 r:2.5 \linewd 0.5  \lcir r:2.5
\linewd 0.1
\rlvec(13 13)
\htext(144 1.5){\large${{\mathcal P}^*_{ \lambda}}$}
\move(223  1) \linewd 0.1
  \rlvec(-13 -1)
\htext(224 -1){\large${{\mathcal P}^*_{ \mathrm{CEP}}}$}
\htext(209.5 21.5){\large ${ y_h}$}
\htext(197.5 31){\large${ y_t}$}
\htext(182 -14.5){\large${ y=d}$}
\htext(126 12){\large${ y_h}$}
\htext(139.5 21){\large${ y_t}$}
\move (210 0 ) \linewd 1
\arrowheadtype t:F \ravec(-22 -8.5)
\end{texdraw}
\end{center}
\caption{Schematic RG flow pattern found for critical end points in position-space RG calculations \cite{BW76,KGYF81}. The fixed point
${\mathcal{P}}^*_\mathrm{CEP}$, representing the critical end point, has two relevant
eigenperturbations whose eigenexponents match those of the fixed point
${\mathcal{P}}^*_\mathrm{\lambda}$, and a discontinuity eigenperturbation
along the trajectory connecting the two fixed points, with
eigenexponent $y=d$. }
\label{fig:RGflpat}    
\end{figure}

Second, the above RG scenario for critical end points
clearly differs from the one encountered in position-space RG analyses of
lattice models \cite{BW76,KGYF81}. These yielded the RG flow
pattern depicted in Fig.~\ref{fig:RGflpat}, in which the critical line is mapped
onto the fixed point ${\mathcal{P}}^*_\mathrm{\lambda}$, while the critical end point
is described by a separate fixed point ${\mathcal{P}}^*_\mathrm{CEP}$.
Since separate critical fixed points normally represent
{\em distinct universality classes\/}
of critical behavior, it would be natural to expect that
the corresponding critical RG eigenexponents $y_t=1/\nu$ and $y_h=\Delta/\nu$ 
take on {\em different values\/} at these fixed points. In Ref.\ \cite{BW76,KGYF81}
these values were found to match. Further, ${\mathcal{P}}^*_\mathrm{CEP}$ was found
to involve an additional relevant eigenvector
with eigenexponent $y=d$, characteristic of  a discontinuity fixed point \cite{NN75}.

Whether this RG scenario---two separate fixed points
${\mathcal{P}}^*_\mathrm{\lambda}$ and ${\mathcal{P}}^*_\mathrm{CEP}$
with matching critical spectra---is
correct, remains to be seen. According to Fisher and Barbosa \cite{FB91},
it need not be the invariable rule, even though it should normally be expected. 
If the latter is true, then one ought to be able to corroborate this RG scenario
by means of a systematic derivation based on  the field-theoretic RG approach.
This should also yield a
discontinuity eigenexponent at  ${\mathcal{P}}^*_\mathrm{CEP}$,
which the field-theoretic analysis of Ziman {\em et al.\/} \cite{ZAGJ82}
was unable to give.

A natural way to tackle this problem is
 to consider models involving \emph{two} fluctuating
densities, namely, a (primary) order parameter field $\phi$
and a secondary (noncritical) density $\psi$.
The reason should be clear:
Across $\boldsymbol{\rho}$, the order parameter
has a jump singularity, but
across $\boldsymbol{\sigma}$, another density---the
thermodynamic density conjugate to $g$---is discontinuous.
Obviously, a proper thermodynamic description requires
{\em two\/} thermodynamic densities,
in addition to the thermodynamic field $T$.
Models of this sort have been investigated by means of
Landau and density functional theory \cite{RCC92,DL89b,GD93,GD98},
and by Monte Carlo calculations \cite{WSN98}.
Unless long-range interactions have been included also,
they may be viewed as continuum versions of the
Blume-Emery-Griffiths (BEG) model \cite{BEG71}.

The continuum models with short-range interactions we
shall consider here are of this kind but more general than
the previously studied ones. They have been chosen
in such a way that they can also be used to investigate
interfacial critical phenomena such as the critical adsorption
of one of the two components of the fluid at
the interface between the liquid $\beta\gamma$ phase and the
spectator phase $\alpha$ \cite{FdG78,LF89,Die94a,FD95}.

In the present paper%
\footnote{A brief report of parts of the results presented here
has been given in Ref.\ \cite{DS00a}; see also Ref.\ \cite{Dr-MS99}.}
we shall confine our attention to
\emph{bulk} properties, leaving the study of critical adsorption,
with the challenging question of whether the
spectator phase $\alpha$ may be replaced
by a  hard wall (the `wall assumption'  of Ref.\ \cite{FU90b},
implicit in virtually
all work on critical adsorption at interfaces),
to a forthcoming paper \cite{SD99}.

Our principal aims are the following. First, using field-theoretic
means, we wish to give a systematic derivation
of the coexistence singularities
(\ref{singpb}) and (\ref{singdens})
\emph{beyond the level of Landau theory.}
Our second goal is to show that
the critical behavior at the critical end point
is the same as along the critical line $\lambda$,
inasmuch as universal properties are concerned.
To this end we shall have to clarify whether
the RG flow pattern found in position-space
RG calculation \cite{BW76,KGYF81} prevails in
our field-theoretic analysis.
In particular, we shall verify the appearance
of a $g$-dependent
scaling field with RG eigenexponent $y=d$, and explain
its origin and significance (cf.\ Ref.\ \cite{DS00a}).

An outline of the paper is as follows.
In the next section, we introduce appropriate continuum models
both for the case of symmetric and of
nonsymmetric binary fluids; we
justify their choice on the basis of phenomenological arguments, and
show that they can be obtained from the BEG model by means
of a Kac-Hubbard-Stratonovich transformation and subsequent
continuum approximation. In Section \ref{sec:LT}, we utilize
 the Landau approximation to analyze these models; we verify that
for  suitable choices of their parameters,
phase diagrams with the
correct qualitative features result, determine the Landau-theory analogs
of the coexistence singularities (\ref{singpb}) and (\ref{singdens}),
and work out a number of perturbative results
that will be needed in later sections.
The field-theoretical analysis beyond Landau theory is
taken  up in Section \ref{sec:beyLT}.
Section \ref{sec:concl} summarizes our main results
and conclusions. Finally, the Appendix
contains some details of the mapping of the BEG
model to the employed continuum models.

\section{Models}\label{sec:models}

\subsection{Phenomenological considerations}\label{sec:models1}

The models we wish to consider have a Hamiltonian of the form
\begin{equation}\label{ham}
{\mathcal{H}}[\phi,\psi]={\mathcal{H}}_1[\phi]+{\mathcal{H}}_2[\psi]
+{\mathcal{H}}_{12}[\phi,\psi]\;,
\end{equation}
where $\phi(\boldsymbol{x})$ and $\psi(\boldsymbol{x})$ are scalar fields
on a  bounded region $\Omega$
of the $d$-dimensional Euclidean space ${\mathbb R}^d$.
It is understood that the limit $\Omega \uparrow {\mathbb R}^d$ is taken;
since we shall only be concerned  with bulk properties in the present paper,
one can imagine $\Omega$ to be a $d$-dimensional hypercube
with periodic boundary conditions whose linear extension is infinite.
While $\phi$, the primary order parameter field, becomes critical, the
secondary field, $\psi$, is noncritical.%
\footnote{The term `noncritical'
is not meant to
imply that the correlation functions of the $\psi$
field do not display any critical singularities.
Owing to the coupling between between $\psi$ and $\phi$, correlation
functions of the $\psi$ field also display critical behavior,
as should be clear and will
be shown explicitly below.  Only if this coupling
vanishes,  are the correlation functions of $\psi$
those of a massive field theory.}

To simplify matters, we shall ignore long-range interactions.
In real binary fluids, long-range interactions of the van der Waals' type
are normally present. While these play a crucial role for ---
and would have to be included in a study of ---
wetting phenomena \cite{Die88}, they can be trusted to be irrelevant
in the RG sense for critical phenomena. Anticipating that a gradient
expansion be made, we presume that the Hamiltonian is a local functional
depending on $\phi$, $\psi$, and their gradients.

Consider first the case of
a hypothetical binary fluid whose properties are {\em symmetric\/} with
regard to interchange of the two constituents $\varsigma = $ A, B.
If  the interaction between  $\varsigma$ and $\varsigma'$
particles is
described by a pair potential $U_{\varsigma\varsigma'}(\boldsymbol{r})$,
then this symmetry is realized if $U_{\mathrm{AA}}=U_{\mathrm{BB}}$
and $\mu_\mathrm{A}=\mu_\mathrm{B}$, where $\mu_{\mathrm{A,B}}$ are the respective
chemical potentials.
For such symmetric binary  fluids, one can identify $\phi$ and $\psi$
(up to convenient normalization factors)  with the difference
$\varrho_{\mathrm{A}}-\varrho_{\mathrm{B}}$
and sum $\varrho_{\mathrm{A}}-\varrho_{\mathrm{B}}$
of the total local mass densities $\varrho_{\mathrm{A,B}}$,
respectively. The Hamiltonian must be even in $\phi$;
that is, ${\mathcal{H}}$, ${\mathcal{H}}_1$,
and ${\mathcal{H}}_{12}$ must satisfy the condition
\begin{equation}\label{Z2symmetry}
{\mathcal{H}}[-\phi,\psi]={\mathcal{H}}[\phi,\psi]\;.
\end{equation}
On the other hand, terms even {\em and\/} odd in $\psi$ are allowed.

We shall keep in ${\mathcal{H}}_1$ monomials up to
fourth order in $\phi$, in
${\mathcal{H}}_2$ those to the same order in $\psi$, and in
${\mathcal{H}}_{12}$ terms up to order $\phi^2\psi$.
Further, contributions of higher than second order
in the gradient operator $\nabla$
will be discarded. It is convenient to make an $\boldsymbol{x}$-independent shift
\begin{equation}\label{psishift}
\psi(\boldsymbol{x})\to \psi(\boldsymbol{x})+\Psi\;,
\end{equation}
choosing $\Psi$ such that the coefficient of the
$\psi^3$ term in ${\mathcal{H}}_2$ vanishes.
The requirement of invariance under the Euclidean group ${\mathbb{E}}(d)$
gives further restrictions, forbidding, in particular, terms linear in $\nabla$
such as ${\int_\Omega}\, (\phi\nabla\psi-\psi\nabla\phi)$. Noting that
${\int_\Omega}\,\psi\triangle\phi^2$ equals ${\int_\Omega}\,\phi^2\triangle\psi$ up to boundary terms,
we arrive at
\begin{equation}\label{H1}
{\mathcal{H}}_1[\phi]={\int_\Omega}\left[
\frac{A}{2}\left(\nabla\phi\right)^2+\frac{a_2}{2}\, \phi^2+
\frac{a_4}{4}\,\phi^4-h\phi
\right],
\end{equation}
\begin{equation}\label{H2}
{\mathcal{H}}_2[\psi]={\int_\Omega}\left[
\frac{B}{2}\left(\nabla\psi\right)^2+\frac{b_2}{2}\, \psi^2+
\frac{b_4}{4}\,\psi^4-g\psi
\right],
\end{equation}
and
\begin{eqnarray}\label{H12}
{\mathcal{H}}_{12}[\phi,\psi]&=&{\int_\Omega} 
\left[d_{11}\,\phi\,\psi+
\frac{d_{21}}{2}\,\phi^2\,\psi+e_{11}\,\phi\,\triangle\psi
\right.\nonumber\\&&\mbox{}+
\frac{e_{21}}{2}\,\phi^2\,\triangle\psi
+\frac{f_{21}}{2}\,{\left(\nabla\phi\right)}^2\,\psi
\biggr]
\end{eqnarray}
with
\begin{equation}\label{oddcoupl}
h=d_{11}=e_{11}=0\;.
\end{equation}
At the Gaussian fixed point at which $\phi$ is critical while  
$\psi$ is noncritical, the coefficient $a_2$
vanishes, but $b_2$ remains nonzero. Since the coefficients $A$ and $b_2$
could be transformed into unity by a change of normalization
of $\phi$ and $\psi$, it is appropriate to take $A$ and $b_2$ as
dimensionless \cite{rem:Ab2}. The momentum dimensions of the other
parameters are listed in Table \ref{tab:cmd}.

\begin{table}
\caption{Canonical momentum dimensions for $d=4-\epsilon$}
\label{tab:cmd}
\begin{tabular}{ccccccccc}
\hline\noalign{\smallskip}
$\phi$&$\psi$&$a_2$ & $a_4$ &$h$&$A$& $B$ \\
\noalign{\smallskip}\hline\noalign{\smallskip}
$1-\frac{\epsilon}{2}$&$2-\frac{\epsilon}{2}$&
$2$&$\epsilon$&$3-\frac{\epsilon}{2}$&$0$&$-2$\\
\noalign{\smallskip}\hline\hline\noalign{\smallskip}
$b_2$&$b_4$&$g$&$d_{11}$&$d_{21}$&$e_{11}$&$e_{21},f_{21}$\\
\noalign{\smallskip}\hline\noalign{\smallskip}
$0$&$\epsilon-4$&$2-\frac{\epsilon}{2}$&$1$&$
\frac{\epsilon}{2}$&$-1$&$\frac{\epsilon}{2}-2$\\
\noalign{\smallskip}\hline\noalign{\smallskip}
\end{tabular}
\end{table}

If the restriction $\mu_{\mathrm{A}}=\mu_{\mathrm{B}}$ is lifted by
turning on a chemical potential difference
$\mu_{-}=\mu_{\mathrm{A}}-\mu_{\mathrm{B}}$ while the AB-symmetry of
the microscopic interactions persists,
then property (\ref{Z2symmetry}) will be lost,
the interaction constants in (\ref{oddcoupl})
will no longer vanish, and a $\phi^3$ term is to be expected in
${\mathcal{H}}_1$. The latter can be transformed away by making
a shift
\begin{equation}\label{phishift}
\phi(\boldsymbol{x})\to \phi(\boldsymbol{x})+\Phi\;,
\end{equation}
analogous to (\ref{psishift}). The coefficients of all
contributions that are odd in $\phi$ must change sign as
$\mu_-$ is reversed. This property carries over to $\Phi$.
Hence these interaction constants as well as $\Phi$ vary $\sim \mu_-$
for small $\mu_-$. 

In the general case, in which the interactions are not AB-symmetric,
the interaction constants of the terms of ${\mathcal{H}}$
that break the $\phi\to-\phi$ symmetry
are no longer odd in $\mu_-$ and hence do not vanish
for $\mu_-=0$. The same applies to the shift $\Phi$. However,
the Hamiltonian
specified by (\ref{ham}) and (\ref{H1})--(\ref{H12})
remains appropriate. The main difference is
that the coefficients $d_{11}$ and $e_{11}$ do not vanish and
we must consider nonzero values of  the ordering field $h$.

Our choice of monomials retained in ${\mathcal{H}}[\phi,\psi]$ requires
some comments. The $\phi$-dependent part
${\mathcal{H}}_1$ is the standard $\phi^4$ Hamiltonian,
comprising all those monomials whose coefficients have nonnegative
momentum dimensions for small $\epsilon=4-d\ge 0$
(cf.\ Table \ref{tab:cmd}) and are not redundant
(as is the $\phi^3$ term) \cite{Weg76}. If besides ${\mathcal{H}}_1$
only the terms quadratic and linear in ${\mathcal{H}}_2$ and
the one $\propto d_{21}$ in ${\mathcal{H}}_{12}$ were kept,
then the Hamiltonian ${\mathcal{H}}$ would agree with the one
utilized in the definition of the familiar dynamic model C
\cite{HHM72}.

The $\psi^4$ term with $b_4>0$ has been included for two reasons.
First, if $b_2<0$, it is needed for stability. Second,
we ought to be able to obtain a spectator phase boundary $\sigma$
from our model. Therefore it is natural to require the model to yield 
such a coexistence surface already
{\em in the absence of coupling to the primary field\/} $\phi$
(i.~e., for ${\mathcal{H}}_{12}=0$). The obvious analog of ${\boldsymbol{\sigma}}$
for this case (and given $b_4>0$) is the plane $g=0$; within Landau
theory, the two (spectator and `liquid') bulk phases coexisting there
for $b_2<0$
correspond to the expectation values
$
\langle\psi\rangle=\pm \sqrt{|b_2|/ b_4}
$.
Note that,
rewritten in terms of the shifted and rescaled field
\begin{equation}
\psi_\pm=\sqrt{|b_2|}\left(\psi\mp\sqrt{|b_2|\over b_4}\right)\;,
\end{equation}
${\mathcal{H}}_2$ takes the form
\begin{eqnarray}\label{H2pm}
{\mathcal{H}}_2[\psi]&=&{\int_\Omega}\left[
\frac{B}{2|b_2|}\left(\nabla{\psi_\pm}\right)^2+ \psi_\pm^2-g\,{\psi_\pm}
\right.\nonumber\\&&\mbox{}\left.
\pm \sqrt{b_4}\,\psi_\pm^3+\frac{b_4}{4}\,
\psi_\pm^4+\frac{1}{4b_4}\mp {g\over\sqrt{b_4}}
\right].
\end{eqnarray}
Thus its $\psi_\pm^2$ term is positive and independent of $b_4$.
The coefficients of the $\psi_\pm^3$ and $\psi_\pm^4$ terms in (\ref{H2pm}) (which have {\em negative\/} momentum dimensions) vanish as $b_4\to 0$.
If we let these coefficients as well as the one $\propto B$ of the
$(\nabla\psi_\pm)^2$ monomial (whose momentum dimension is also negative)
approach zero, and set
the field $g$ to the value $g=\pm 0$,
then the $\psi_\pm$-dependent part of
${\mathcal{H}}_2$ turns into precisely this quadratic part, namely the
Gaussian Hamiltonian
\begin{equation}\label{HG}
{\mathcal{H}}_{\mathrm{G}}[\psi_\pm]={\int_\Omega} \psi_\pm^2\;.
\end{equation}
For vanishing ${\mathcal{H}}_{12}$,
this is a fixed-point Hamiltonian in the space of $\psi$-dependent
Hamiltonians. We refrain from a more detailed discussion of
RG issues here, reserving it to later sections. Note, however, that
when the previously mentioned gradient, cubic, and quartic terms
of ${\mathcal{H}}_2$
are dropped together with all contributions in ${\mathcal{H}}_{12}$
other than the one $\propto d_{21}$, then the total Hamiltonian ${\cal H}$
reduces to the one of dynamic model C. A trivial, but useful,
observation is that the constant part of the Hamiltonian (\ref{H2pm})
has a {\em $g$-dependent\/} part. Any constant in the integrand of
a bulk Hamiltonian couples to
$|\Omega|\equiv {\int_\Omega}$, the volume, and hence scales like a relevant
scaling field with scaling index $d$ under RG transformations.
In Wegner's terminology \cite{Weg76}, such a scaling field is called
{\em special\/} rather than relevant,
because of its well-known relationship with the analytic
part of the bulk free energy density. In any case, we see
already at this stage (where the coupling to the order parameter field $\phi$
has not yet been taken into account) a scaling field emerging
that is odd in $g$ and has scaling index $d$. 

As stressed earlier, $b_4$ must not simply be set to zero:
It must have a positive value in order that our model may yield
a bulk phase diagram with the correct topology, for appropriate
values of the other parameters (cf.\ Sec.\ \ref{sec:LT}).
The coefficient $B$, on the other hand, can be set to zero in essentially
all of our subsequent analysis. Our main reason for including this
term is our intention of employing the model
in a forthcoming paper \cite{SD99} in a study
of critical adsorption of binary fluid mixtures
at their interface with the spectator gas phase.
The coefficient $B$ serves to provide a scale (correlation length)
on which kink solutions $\check{\psi}_{-+}(z)$
of the Ginzburg-Landau equation for ${\mathcal{H}}_2$,
connecting the asymptotic bulk solutions $\mp \sqrt{|b_2|/b_4}$
at $z=\pm\infty$,
vary along the direction perpendicular to the interface.
As $B\to 0$, these kinks become sharp steps.
One does not need to be concerned about the vanishing of this scale here
since we shall only deal with bulk properties below.

Likewise, the Laplacian terms $\propto e_{11}$ and $\propto e_{21}$
as well as the term $\propto f_{21}$
(all of which have {\em negative\/} momentum dimensions)
have been introduced primarily with a view to
our intended analysis of interfacial problems \cite{SD99}, and
can be ignored in the sequel.
However, the coupling term $\propto d_{11}$, which is also
not present in the Hamiltonian of dynamic model C,
must {\em not\/} be ignored
in the general nonsymmetric case. As the reader might anticipate and
will see below, it plays an important role in producing
the  $|{t}|^\beta$ part of the coexistence singularity (\ref{singdens}).

\subsection{Derivation from the Blume-Emery-Griffiths model}
\label{sec:derivBEG}

A familiar lattice model for binary mixtures that exhibits a
critical end point is the BEG model \cite{BEG71,CL95}.
This is a classical spin-$1$ lattice model with the Hamiltonian
\begin{eqnarray}\label{HBEG}
{\mathcal H}_\mathrm{BEG}[\boldsymbol{S}]&=&-\sum_{\langle \boldsymbol{i},\boldsymbol{j}\rangle}
\left[J\,S_{\boldsymbol{i}}\,S_{\boldsymbol{j}}
+K\,S_{\boldsymbol{i}}^2\,S_{\boldsymbol{j}}^2
\right.\nonumber\\&&\left.\quad\qquad\mbox{}+L\left(
S_{\boldsymbol{i}}^2\,S_{\boldsymbol{j}}
+S_{\boldsymbol{i}}\,S_{\boldsymbol{j}}^2\right)\right]
\nonumber\\
&&\mbox{}-\sum_{\boldsymbol{i}}\left(H\,S_{\boldsymbol{i}}+D\,S_{\boldsymbol{i}}^2 \right),
\quad S_{\boldsymbol{i}}=0,\pm 1\,,
\end{eqnarray}
where 
$\langle \boldsymbol{i},\boldsymbol{j}\rangle$ indicates summation over
nearest-neighbor
pairs of lattice sites.
Initially it was proposed to simulate
$^3$He-$^4$He mixtures; in this interpretation,
a $^3$He atom at site $\boldsymbol{i}$ corresponds
to $S_{\boldsymbol{i}}=0$ and a
$^4$He atom to $S_{\boldsymbol{i}}=\pm 1$  \cite{BEG71},
so there is one and only one helium
atom at each lattice site, and
the model makes no allowance for vacancies.

However, the model may also be interpreted in a distinct---%
and for our purposes more appealing---fashion as
a lattice-gas model for a binary fluid \cite{SL75a}.
Then $\boldsymbol{i}$ is a label for microscopic cells that
can hold at most a {\em single\/} A or B particle (atom or molecule),
and the states $S_{\boldsymbol{i}}=+1$, $-1$, and $0$ correspond,
respectively,
to the cases `cell $\boldsymbol{i}$ is occupied by an A
particle', `cell $\boldsymbol{i}$ is occupied by a B
particle', and `cell $\boldsymbol{i}$ is empty'.
Thus the variables $H$ and $D$
represent odd and even linear combinations
of the chemical potentials
$\mu_\mathrm{A}$ and $\mu_\mathrm{B}$ \cite{SL75a}:
\begin{equation}
H={\mu_\mathrm{A}-\mu_\mathrm{B}\over 2\,k_{\mathrm{B}}T}
\end{equation}
and
\begin{equation}
D={\mu_\mathrm{A}+\mu_\mathrm{B}\over 2\,k_{\mathrm{B}}T}\;.
\end{equation}
Without loss of generality, we can take $L>0$ \cite{SL75a}.
We also assume that  $J>0$ and $K>0$. Thus,
 as $J$ increases,
the tendency for phase separation in the liquid state grows,
and larger values of $K$ correspond to a stronger drive
of condensation of the mixture. 

On sufficiently large length scales, the physics of the BEG model
should be described by the continuum models introduced in the previous
subsection. Hence
one ought to be able to derive the latter from it.
This may be achieved in much the same way as
one can map
the Ising model onto a continuum field theory, utilizing a
Kac-Hubbard-Stratonovich transformation (see, e.g., Ref.~\cite{Fis83a}).
Details are given in the Appendix, where we show that the
grand partition function of the BEG model,
\begin{equation}
{\mathcal{Z}}_\mathrm{BEG}=\sum_{\{S_{\boldsymbol{i}}=0,\pm1\}}
\mathrm{e}^{-{\mathcal{H}}_\mathrm{BEG}\vec[S]}\;,
\end{equation}
can we rewritten exactly as an integral over continuous fields
$\phi_{\boldsymbol{i}}$ and $\psi_{\boldsymbol{i}}$, that is to say, as
the partition function of a \emph{lattice field theory}. One has
\begin{equation}\label{ZBEG}
{\mathcal{Z}}_\mathrm{BEG}=\mathrm{e}^{-f_0(J,L,K)}
\left[\prod_{\boldsymbol{i}}{\int_{-\infty}^{+\infty}}\!
d\phi_{\boldsymbol{i}}{\int_{-\infty}^{+\infty}}\!d\psi_{\boldsymbol{i}}\right]
\mathrm{e}^{-{\mathcal{H}}_\mathrm{lft}[\boldsymbol{\phi},\boldsymbol{\psi}]}\:.
\end{equation}
The explicit form of this Hamiltonian ${\mathcal{H}}_\mathrm{lft}[\boldsymbol{\phi},\boldsymbol{\psi}]$
is given in (\ref{Hlft}) of the Appendix. As is shown there,
making a continuum approximation then yields a Hamiltonian
of the form specified
by (\ref{ham}) and (\ref{H1})--(\ref{H12}).

\section{Landau theory}\label{sec:LT}

Landau theory is of value not only because of its simplicity
and ability to reproduce essential topological features of
the phase diagram, but also because it serves as the starting point
of studies based on RG improved perturbation theory.
Our main goal here is to convince ourselves that our model
indeed yields  a bulk phase diagram
with a critical end point
and the topology illustrated in Fig.~\ref{fig:scpd},
providing the model parameters are in the appropriate range.

Since we shall only consider translationally invariant states here,
we can take $\phi$ and $\psi$ to be position-independent
in the rest of this section. Hence all terms of ${\mathcal{H}}$
involving spatial derivatives do not contribute.
If we drop these terms, then its part even in $\phi$
becomes equivalent to the Hamiltonian investigated in Ref.\ \cite{RCC92}
via Landau theory. 
Thus, for the symmetric case, the results may be partly inferred
from this reference.

In the Landau approximation, the grand potential
(per volume and $k_{\mathrm{B}}T$)
is given by
\begin{equation}
{\mathcal{A}}(a,b,d,h,g) = \inf_{\phi,\psi} 
{ \mathcal{V}} (\phi,\psi) \;, 
\label{Landau.Jgk}
\end{equation}
with
\begin{equation}
{\mathcal{V}} (\phi,\psi) = \frac{1}{|\Omega|}
{\mathcal{H}}[\phi,\psi] \;,
\label{Landaupot}
\end{equation}
where $a$, $b$, and $d$ stand for the sets of parameters $\{a_2,a_4\}$,
$\{b_2,b_4\}$, and $\{d_{11},d_{21}\}$, respectively.
From (\ref{H1})--(\ref{H12}) we find 
\begin{eqnarray}
{\mathcal{V}}(\phi,\psi)&=&\frac{1}{2}\,\left(a_2+d_{21}\,\psi\right) \phi^2+
\frac{a_4}{4}\,\phi^4-h \phi\nonumber\\&&\mbox{}
+\frac{b_2}{2}\, \psi^2+
\frac{b_4}{4}\,\psi^4+d_{11}\,\psi\phi- g \psi\;.
\end{eqnarray}
For the sake of thermodynamic stability,
$a_4$ and $b_4$ are assumed to be positive.

In place of $g$ we shall occasionally use the conjugate
density
\begin{equation}\label{psicheck}
\check{\psi}\equiv -\left.\frac{\partial{\mathcal{A}}}{\partial g}
\right|_{a,b,d,h}
\end{equation}
as independent
thermodynamic variable, utilizing the mixed field-density representation
$(a,b,d,h,\check\psi)$ instead of the field representation
$(a,b,d,h,g)$.
The thermodynamic potential associated with the former
is defined by
\begin{equation}\label{mixedpot}
{\mathcal{B}}(a,b,d,h,\check\psi)=
\inf_\phi\left\{{\mathcal{V}}(\phi,\check\psi)+g\check\psi\right\}.
\end{equation}

\subsection{Symmetric case}\label{sec:LTSymc}

We first consider the symmetric case, setting $d_{11}=0$.
Owing to the implied invariance with regard to $(\phi,h)\to (-\phi,-h)$,
we may restrict ourselves to values $h>0$. Since a sign change of $d_{21}$
can be compensated by simultaneous sign changes of $\psi$ and $g$,
we may furthermore presume that
\begin{equation}\label{d21ineq}
d_{21}\le 0\;.
\end{equation}
From the analysis
presented in Ref.\ \cite{RCC92} it is clear that a variety
of different types of phase diagrams (with critical points,
critical end points, tricritical points, or special tricritical points)
can be obtained, depending on the chosen range of
the parameters $a_2$, $b_2$, $d_{21}$, and $g$.
Our aim here is not to explore all these possibilities; rather we shall
focus on the case of critical end points, and choose the
values of these parameters accordingly.

The equilibrium densities must satisfy the (classical) equations of state
\begin{equation}\label{cleqstgen}
\frac{\partial{\mathcal{V}}(\phi,\psi)}{\partial\phi}=
\frac{\partial{\mathcal{V}}(\phi,\psi)}{\partial\psi}=0\;,
\end{equation}
which in the present symmetric case become 
\begin{eqnarray}
\label{symm.GLEphi}
\left(a_2+d_{21}\, \psi\right) \phi + a_4\, \phi^3 &=& h\;,\\[0.5em]
\label{symm.GLEpsi}
b_2\,\psi + b_4\,\psi^3 +\frac{d_{21}}{2}\,\phi^2&=& g\;.
\end{eqnarray}
From the $\phi\to -\phi$ symmetry (\ref{Z2symmetry}) it is clear that
the $\beta\gamma$ coexistence surface $\boldsymbol{\sigma}$
of the ordered ($\phi\ne 0$) states must lie in the $h=0$ plane.
Upon setting $h=0$, the first equation of state, (\ref{symm.GLEpsi}),
can easily be solved for
$\phi$ to determine that value $\phi_{\mathrm{min}}(\psi)$
at which ${\mathcal{V}}(\phi,\psi)$ becomes minimal for the given value of
$\psi$. One finds
\begin{equation}\label{phimin}
\phi_{\mathrm{min}}(\psi)=\cases{0\,,&if $a_2 +d_{21}\,\psi >0$.\cr\cr
\pm\sqrt{|a_2+d_{21}\,\psi|\over a_4}\,,&if $a_2 +d_{21}\,\psi <0$.}
\end{equation}
The critical value of $\psi$ at which the bifurcation occurs,
\begin{equation}
\psi_{\lambda}= -\frac{a_2}{d_{21}}\;,
\end{equation}
is the solution to
\begin{equation}
\frac{\partial^2}{\partial \phi^2}\,
{\mathcal{V}}(0,\psi)=0\;,
\end{equation}
a condition that must be fulfilled whenever $\phi$
is critical at $\phi=0$ and hence on the $\lambda$-line.

Upon inserting (\ref{phimin}) into
the right-hand side of (\ref{mixedpot}), we obtain
\begin{eqnarray}
{\mathcal{B}}(a,b,c,0,{\psi})&=&
\frac{b_2}{2}\,{\psi}^2 + 
  \frac{b_4}{4}\,{\psi}^4
\nonumber\\&&\mbox{}
- \theta{\left(a_2\,{\psi-\psi_{\lambda}\over \psi_{\lambda}} \right)}\,
 \frac{{d_{21}^2}}{4\,a_4}\,\left({\psi}-\psi_{\lambda} \right)^2\,,\quad
\end{eqnarray}
where $\theta(.)$ is the step function.

Since we wish the density $\psi$ to be positive on
the $\lambda$-line we take
\begin{equation}\label{a2ineq}
a_2\ge 0\;;
\end{equation}
together with the inequality (\ref{d21ineq}) this ensures that
$\psi_{\lambda} >0$. Under these conditions the trivial solution
$\phi_{\mathrm{min}}=0$ holds in the regime $\psi<\psi_{\lambda}$.
Thus the equilibrium densities $\psi_\alpha$ and $\psi_{\beta\gamma}$
of the disordered vapor and liquid phases are solutions to the $\phi=0$
analog of (\ref{symm.GLEpsi}):
\begin{equation}\label{symmdis.GLEpsi}
b_2\,\psi + b_4\,\psi^3 = g\;.
\end{equation}
If $b_2>0$, there exists a unique real solution, which
is positive or negative, depending on whether $g>0$ or $g<0$,
namely
\begin{eqnarray}\label{eta0}
\eta_0(b_2,b_4,g)&=&\frac{1}{3}\,(r_0)^{1/3}
-\frac{b_2}{b_4}\,(r_0)^{-1/3}
\label{psi0}
\end{eqnarray}
with
\begin{equation}\label{r0}
r_0=\frac{27}{2}\,\frac{g}{b_4}+
\sqrt{\frac{729}{4}\left(\frac{g}{b_4}\right)^2 + 
              27\,
\left(\frac{b_2}{b_4}\right)^3}\;.
\end{equation}
It has the power-series expansion
\begin{eqnarray}\label{eta0exp}
\eta_0(b_2\ge 0,b_4,g)={\frac{g}{b_2}} - 
  {\frac{b_4\,{g^3}}{{b_2^4}}} + 
  {\frac{3\,{b_4^2}\,{g^5}}
    {{b_2^7}}} + {{O}}\big(g^7\big)\;,
\end{eqnarray}
where the subscript $0$ is to remind us that
the solution vanishes for $g=0$.
For $b_2<0$, (\ref{symmdis.GLEpsi}) has three real solutions
if
\begin{equation}
|g|< g_<\equiv \frac{2\,|b_2|^{3/2}}{3\,\sqrt{3\,b_4}}\;,
\end{equation}
and a single real
one if $|g|>g_<$. We denote the ones that turn
in the limit $g\to 0$ into the nontrivial
$g\ne 0$ solutions $\pm \sqrt{|b_2|/b_4}$ as $\eta_\pm(b_2,b_4,g)$.
They have the property
\begin{equation}\label{+-sym}
\eta_-(b_2,b_4,g)=-\eta_+(b_2,b_4,-g)
\end{equation}
and the power-series expansion
\begin{eqnarray}\label{etapmexp}
\eta_\pm(b_2\le 0,b_4,g)&=&\pm
\sqrt{\frac{|b_2|}{b_4}} + 
  {\frac{g}{2\,|b_2|}} \mp {\frac{3\,{\sqrt{{b_4}}}\,
      {g^2}}{8\,{|b_2|^{{{5}/{2}}}}}}
 \nonumber\\&&\mbox{}
+ {\frac{{b_4}\,{g^3}}{2\,{|b_2|^4}}}\mp
  {\frac{105\,{b_4^{{{3}/{2}}}}\,{g^4}}
    {128\,{|b_2|^{{{11}/{2}}}}}}
 \nonumber\\&&\mbox{} 
+{\frac{3\,{b_4^2}\,{g^5}}{2\,{|b_2|^7}}} + 
  {{O}}\big(g^6\big)\;.
\end{eqnarray}
Further, $\eta_+$ is given by the analog
of the result (\ref{psi0}) for $\eta_0$
one obtains through the replacement $b_2\to-|b_2|$, both in
(\ref{psi0}) and the expression (\ref{r0}) for $r_0$.

In terms of these solutions, the
densities of the \emph{disordered fluid state} become
\begin{equation}
\phi=0\,,\;\psi=\eta_0(b_2,b_4,g)\;,\;\mbox{ for } b_2>0\,,\;h=0\,,\;\psi<\psi_{\lambda}\;,
\end{equation}
while the densities of the
\emph{disordered vapor phase} $\alpha$ and
the \emph{disordered liquid phase} $\beta\gamma$
read
\begin{equation}\label{alphabetagamma}
\begin{array}{lcr}(\phi_\alpha,\psi_\alpha)&=& (0,\eta_-(b_2,b_4,g))\\[0.5em]
(\phi_{\beta\gamma},\psi_{\beta\gamma})&=& (0,\eta_+(b_2,b_4,g))
\end{array}\quad \mbox{for }
\left\{\begin{array}{l}b_2<0\,,\\h=0\,,\\\psi<\psi_{\lambda}\;.%}
\end{array}\right.
\end{equation}

In the latter two cases, the solutions $\eta_+$ ($\beta\gamma$ phase)
and $\eta_-$ ($\alpha$ phase) are the thermodynamic stable
ones if $g>0$ and $g<0$, respectively;
the remaining two real roots of (\ref{psi0}) one has
if $|g|<g_<$ correspond to metastable
and unstable states.

At $g=0$, these states coexist down to that value of $b_2$
at which $\psi_{\beta\gamma}$ intersects the line $\psi=\psi_{\lambda}$,
which is
\begin{equation}
b_{2e}=-b_4\,\psi_{\lambda}^2=-b_4 \left(\frac{a_2 }{d_{21}}\right)^2,
\end{equation}
provided
\begin{equation}\label{condnotcp}
\psi_{\lambda}>\frac{-d_{21}}{2\,\sqrt{a_4\,b_4}}\;.
\end{equation}
As we shall show below, this condition guarantees that no tricritical
point appears. The resulting phase diagram is displayed in Fig.~\ref{fig:pdsmixed} in a mixed field/density representation,
and in Fig.~\ref{fig:pds} in a field representation; it
corresponds to case (e)
of Ref.\ \cite{RCC92}.%
\footnote{In order to check the consistency
with Ref.\ \cite{RCC92}, one
should identify the variables $\varrho$, $a$, $\mu$
and $A$ utilized there 
with $-\psi$, $a_2$, $-g$, and $b_2$, respectively,
and set $a_4=d_{21}=b_2=b_4=1$.} 
There is a liquid-vapor critical point at
$b_2=\psi=0$ and a critical end point
at $b_2=b_{2e}$, $\psi=\psi_{\lambda}$.
The boundaries of the $\alpha$-$\beta\gamma$ coexistence region
for $b_{2e}<b_2<0$ are given by the $g=0$ solutions $\eta_\pm$
of (\ref{etapmexp}). 

\begin{figure}[htb]
%
%The figure is generated here via texdraw%%%
\begin{center}
\begin{texdraw}
\drawdim mm \setunitscale 1 
\move (0 0) %\fcir f:0 r:0.75
\move(-30 0) \arrowheadtype t:W \arrowheadsize l:3 w:2 \ravec(70 0)
\htext(32 -5){$\psi/\psi_{\lambda}$}
\move(0 0) \ravec(0 20) \htext(-4 21){$b_2/|b_{2e}|$}
\htext(-15 -14){\begin{minipage}[b]{30mm}\begin{center}%
coexistence\\ of $\alpha$ and $\beta\gamma$\end{center}\end{minipage}}
\htext(-20 -25){\begin{minipage}[b]{40mm}\begin{center}%
coexistence of $\alpha$, $\beta$, and $\gamma$\end{center}\end{minipage}}
\linewd 0.75
%parabola part of phase boundary%
\move(20 -20)
\clvec(19 -18.05)(18 -16.2)(17 -14.45)
\clvec(16 -12.8)(15 -11.25)(14 -9.8)
\clvec(13 -8.45)(12 -7.2)(11 -6.05)
\clvec(10 -5)(9 -4.05)(8 -3.2)
\clvec(7 -2.45)(6 -1.8)(5 -1.25)
\clvec(4 -0.8)(3 -0.45)(2 -0.2)
\clvec(1 -0.05)(0 0)(-1 -0.05)
\clvec(-2 -0.2)(-3 -0.45)(-4 -0.8)
\clvec(-5 -1.25)(-6 -1.8)(-7 -2.45)
\clvec(-8 -3.2)(-9 -4.05)(-10 -5)
\clvec(-11 -6.05)(-12 -7.2)(-13 -8.45)
\clvec(-14 -9.8)(-15 -11.25)(-16 -12.8)
\clvec(-17 -14.45)(-18 -16.2)(-19 -18.05)
\clvec(-19.5 -19.0125)(-19.75 -19.5031)(-20 -20)
%end of parabola part of phase boundary%
\lfill f:0.9 \linewd 0.2 \lpatt(2 2) \lvec(20 -20)
\lpatt()\linewd 0.75
%start nontrivial part of phase boundary%
\move(22.8512 -26)
 \lpatt() \clvec(22.4027 -25)(21.9445 -24)(21.476 -23)
\clvec(20.9963 -22)(20.5046 -21)(20 -20) \lvec(-20 -20)
\clvec(-20.494 -21)(-20.9764 -22)(-21.4481 -23)
\clvec(-21.9097 -24)(-22.3618 -25)(-22.805 -26)
\ifill f:0.75
\move(-20 -20) \clvec(-20.494 -21)(-20.9764 -22)(-21.4481 -23)
\clvec(-21.9097 -24)(-22.3618 -25)(-22.805 -26)
\move(20 20) \clvec(20 10)(20 0)(20 -20)
\clvec(20.5046 -21)(20.9963 -22)(21.476 -23)
\clvec(21.9445 -24)(22.4027 -25)(22.8512 -26)
\move(-20 -20) \linewd 0.2 \lpatt(0.25 0.5) \lvec(-20 -1)
\lpatt() \move(-20 -0.5)
\lvec(-20 0.5) \move(10 -0.5) \lvec(10 0.5)
\htext(8 1){$0.5$}
\move(-0.5 5) \lvec(0.5 5) \htext(1 4){$0.25$}
\move(-0.5 10) \lvec(0.5 10) \htext(1 9){$0.50$}
\htext(-22 1){$-1$} \htext(-13 1){$-0.5$}
\move(-10 -0.5) \lvec(-10 0.5) \htext(21 -20){CEP}
\move(20 -20) \fcir f:0 r:0.75
\vtext(19 5){$\lambda$-line}
\htext(-17.5 12.5){$h=0$}
\end{texdraw}
\end{center}
\caption{Phase diagram of the symmetric model in
the Landau approximation. The diagram is shown in a mixed field/density
representation and corresponds to the following choice of parameter values:
$a_2=5$, $a_4=b_4=1/6$, $d_{21}=-1/2$, $d_{11}=0$, giving
$\psi_{\lambda}=10$ and $b_{2e}=50/3$.}
\label{fig:pdsmixed}      
\end{figure}\bigskip

\begin{figure}[hbt]
\resizebox{\columnwidth}{!}{%
  \includegraphics{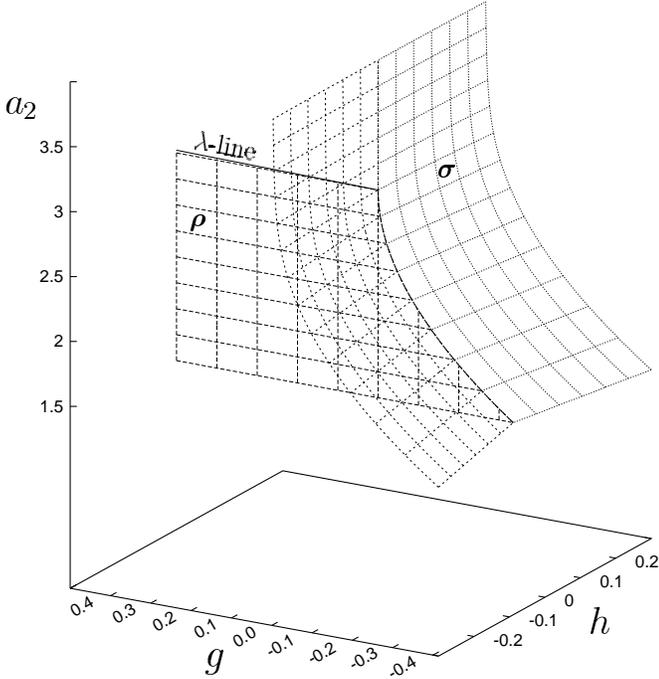}
}
\caption{The analog of Fig.\ \ref{fig:pdsmixed}
in an $a_2gh$ field representation, as obtained by
solving the equations of Landau theory
through numerical means. For $b_2$ we have chosen
the value $b_2=-6$; the remaining parameters
have the same values as in Fig.\ \ref{fig:pdsmixed}, i.e.,
 $a_4=b_4=1/6$, $d_{21}=-1/2$, and $d_{11}=0$.
Only the neighborhood
of the critical end point is shown.}
\label{fig:pds}       
\end{figure}

In order to determine the boundaries of the region of coexistence of the
$\alpha$, $\beta$, and $\gamma$ phases, one generally must resort to
numerical methods. However, for $b_2\lesssim  b_{2e}$, the value $g$
takes on the triple line, $g_\sigma$, is small,
as is the order parameter $\phi$
for $\psi\gtrsim\psi_{\lambda}$ in the vicinity
of the critical end point. This enables us to determine the phase boundaries
in this regime in a perturbative manner. Note, first, that the equilibrium densities
$\psi_\beta$ and $\psi_\gamma$ ($=\psi_\beta$ for $h=0$)
can be written in terms of
\begin{equation}\label{psi+g}
\psi_+(g)\equiv \eta_+(b_2,b_4,g)
\end{equation}
 and expanded
as
\begin{eqnarray}
\psi_{\beta,\gamma}&=&
\eta_+{\left(b_2,b_4,g-\frac{1}{2}\,d_{21}\,%
\phi_{\beta,\gamma}^2\right)}\label{psiphimin}\\&=&
\psi_+(g)-\psi_+{'}(g)\,
\frac{d_{21}}{2}\,\phi_{\beta,\gamma}^2\nonumber\\
&&\mbox{}+\psi_+{''}(g)\,
\frac{d_{21}^2}{8}\,\phi_{\beta,\gamma}^4
+{O}{\left(\phi_{\beta,\gamma}^6\right)}\,,
\label{psi+phiexp}
\end{eqnarray}
with
\begin{equation}\label{psi+prime}
{\psi_+}'(g)=\left[3\,b_4\,\psi_+^2(g)-|b_2|\right]^{-1}
\end{equation}
and
\begin{eqnarray}\label{psi+dprime}
{\psi_+}{''}(g)&=&
-\frac{6\,b_4\,\psi_+(g)}{\left[3\,b_4\,
\psi_+^2(g)-|b_2|\right]^3}\;.
\end{eqnarray}
Inserting the expansion (\ref{psi+phiexp})
into the equation of state (\ref{symm.GLEphi}) with $h=0$ gives
\begin{eqnarray}\label{effeqofst}
\left[a_2+d_{21}\, \psi_+(g)\right] \phi + \left[a_4-\frac{d_{21}^2}{2}\,{\psi_+}'(g)\right] \phi^3\qquad&&\nonumber\\ 
\mbox{}+\frac{d_{21}^3}{8}\,{\psi_+}{''}(g)\,
\phi^5={O}\left(\phi^7\right)\,.&&
\end{eqnarray}
We seek nontrivial solutions of this equation for
\begin{equation}
\check{a}_2{\left[\psi_+(g)\right]} \equiv a_2+d_{21}\,
\psi_+(g) <0\;.
\end{equation}
If the coefficient of the $\phi^3$ term in (\ref{effeqofst}) is positive,
\begin{equation}\label{a4eff}
\check{a}_4{\left[\psi_+(g)\right]} \equiv a_4-\frac{d_{21}^2}{2}\,{\psi_+}'(g) >0\;,
\end{equation}
then the equation (\ref{effeqofst}) for $\phi$ can be solved
in a straightforward manner
 to obtain
\begin{equation}
\phi_\beta=-\phi_\gamma=
\sqrt{\frac{\left|\check{a}_2{\left[\psi_+(g)\right]}\right|}
{\check{a}_4{\left[\psi_+(g)\right]}}}\;%
\left(1+o{\left[\check{a}_2{\left(\psi_+\right)}\right]}\right)\,,
\end{equation}
giving
\begin{equation}\label{psibetasym}
\psi_{\beta,\gamma}\approx\psi_+(g)-
{\psi_+}'(g)\,
\frac{d_{21}}{2}\,
\frac{\left|\check{a}_2{
\left[\psi_+(g)\right]}\right|}
{\check{a}_4{\left[\psi_+(g)\right]}}\;.
\end{equation}

The above analysis suggests that $\check{a}_4[\psi_+(g)]$ plays the
role of an effective $\phi^4$ interaction constant. To verify this
one can calculate
${d^4 {\mathcal{V}}\big[\phi,\psi_{\beta,\gamma}%
(\phi)\big]}/{d \phi^4}\big|_{\phi=0}$
along a path $\psi_{\beta,\gamma}(\phi)$, where
$\psi_{\beta,\gamma}(\phi)$ is given by (\ref{psiphimin}),
with the replacement $\phi_{\beta,\gamma}\to \phi$.
A straightforward calculation gives
\begin{eqnarray}\label{a4effderiv}
\left.\frac{d^4 {\mathcal{V}}{\left[\phi,\psi^{(\beta,\gamma)}%
_{\mathrm{min}}(\phi)\right]}}{d \phi^4}
\right|_{\phi=0}=6\,\check{a}_4{\left[\psi_+(g)\right]}\,,
\end{eqnarray}
where the equation of state (\ref{symm.GLEphi}) and the results
(\ref{psi+prime}) and (\ref{psi+dprime}) for the
derivatives of $\psi_+(g)$ were used.
Hence (\ref{a4eff}) is
the usual condition for the absence of a tricritical point at $\phi=0$.
At the critical end point it becomes
\begin{equation}\label{a4cond}
a_4>\frac{d_{21}^2}{4\,|b_{2e}|}=\frac{1}{b_4}
\left(\frac{d_{21}}{2\,\psi_{\lambda}}\right)^2\,,
\end{equation}
which in turn implies our initially stated condition
(\ref{condnotcp}) if $d_{21}<0$ and
$\psi_{\lambda}>0$, as assumed.

The location of the triple line, $g_\sigma$, can now be determined
from the coexistence condition
\begin{equation}\label{coexcond}
{\mathcal{A}}_\alpha(a,b,d,0,g_\sigma)=
{\mathcal{A}}_{\beta,\gamma}(a,b,d,0,g_\sigma)
\end{equation}
for the grand potentials
\begin{equation}\label{grandpot}
{\mathcal{A}}_{\wp}(a,b,d,h,g)=
{\mathcal{V}}(\phi_\wp,\psi_\wp)\;,\quad\wp=
\alpha,\beta,\gamma\;.
\end{equation}
Upon substituting the above results
for $(\phi_{\beta,\gamma},\psi_{\beta,\gamma})$
together with $(\phi_\alpha,\psi_\alpha)$ from (\ref{alphabetagamma})
into the latter equation, the former one can be
solved beneath the critical end point to obtain
\begin{eqnarray}\label{gsigmasym}
g_\sigma &=&-\left\{\frac{\left[\check{a}_2(\psi)\right]^2}%
{8\,\check{a}_4(\psi)\,\psi}+{O}{\left(
\left[\check{a}_2(\psi)\right]^3\right)}\right\}_{\psi=\psi_+(0)}\\
&=&\mbox{}-
\frac{\left(a_2+d_{21}\,\sqrt{\frac{|b_2|}{b_4}}\right)^2}%
{8\,\left(a_4-\frac{d_{21}^2}{4\,|b_2|}\right)\sqrt{\frac{|b_2|}{b_4}}}
+{O}{\left(\check{a}_{2}^3\right)}
\end{eqnarray}
for $\check{a}_2\big[\sqrt{|b_2|/b_4}\big]=%
a_2+d_{21}\,\sqrt{|b_2|\over b_4}\lesssim 0$.

The latter quantity varies as
$\check{a}_2\approx a_{2,\pm}^0\,|t|$ near the critical end point, while the 
denominator in (\ref{gsigmasym})  approaches a nonvanishing
constant there. 
Remembering that $g_\sigma=0$ for $\check{a}_2>0$ (i.e., $t>0$),
we arrive at
\begin{equation}\label{sndtderivsigma}
\left.\frac{\partial^2g_\sigma(t)}{\partial t^2}\right|_{t=0\pm}=
\cases{0& for $t=0^+$,%
\cr
\frac{-\big(a_{2,-}^0\big)^2}{4\,%
\check{a}_{4e}\,\sqrt{{|b_{2e}|/ b_{4e}}}}&for $t=0^-$,}
\end{equation}
where the subscript $e$ denotes values at the critical end point.
Hence $\partial^2g_\sigma/\partial t^2$ has indeed
a jump singularity at the end point,
as found in other  analyses
of the mean-field type
\cite{RCC92,DL89b,GD93,GD98} and is in conformity with
the $\alpha=0$ analog of the predicted singularity (\ref{singpb}).

Likewise,  the singularity\footnote{%
To obtain the leading temperature singularity given here,
one may replace $g_\sigma(t)$ by its limiting value $g_e=g_\sigma(t=0)$;
the $|t|^2$ singularity (\ref{sndtderivsigma}) of $g_\sigma(t)$
produces {\em subleading\/} nonanalytic
contributions to $\psi_{\beta\gamma/\beta,\gamma}$.}
of the liquid ($\beta\gamma$ or $\beta,\gamma$) branch of the
$\alpha$-$\beta\gamma$ and $\alpha$-$\beta$-$\gamma$
coexistence regions,
\begin{equation}\label{psiliqucs}
\psi_{\beta\gamma/\beta,\gamma}^{\mathrm{sing}}\approx -\frac{d_{21,e}\,|a_{2,-}^0|}{4\,|b_{2e}|\,
\check{a}_{4e}}\,|t|\,\theta(-t)\;,
\end{equation}
resulting from the second term in  (\ref{psibetasym})
is the mean-field analog of the one $\sim|t|^{1-\alpha}$
in (\ref{singdens}); it means that in Landau theory the
first temperature derivative
of $\psi_{\beta,\gamma}$ is {\em discontinuous\/}
at the critical end point (cf.\ Refs.\ \cite{Wil97a,Wil97b}).

The curve in which this portion of the coexistence
boundary intersects the $b_2\psi$ plane---i.e.,
the right branch of the liquid phase boundary
in Fig.\ \ref{fig:pdsmixed}, described by the mapping
$b_2\mapsto \psi_{\beta\gamma}$ and $b_2\mapsto\psi_{\beta,\gamma}$
for $0>b_2>b_{2e}$ and $b_{2e}>b_2$, respectively,%
---also
has a discontinuous slope at the end point.
One finds
\begin{equation}
\left.\frac{\partial \psi_{\beta\gamma/\beta,\gamma}}{\partial b_2}
\right|_{b_2=b_{2e}+0}=\frac{-|d_{21}|}{2\,a_{2}\,b_4}
\end{equation}
and
\begin{eqnarray}
\left.\!\frac{\partial \psi_{\beta\gamma/\beta,\gamma}}{\partial b_2}
\right|_{b_2=b_{2e}- 0}\!&=&
\frac{-|d_{21}|}{2\,a_{2}\,b_4}
-\frac{|d_{21}|^5}{2\,a_2\,b_4\,(4\,a_2^2\,a_4\,b_4-|d_{21}|^4)}
\nonumber\\
&=&-\frac{2\,{a_2}\,{a_4}\,|d_{21}|}{4\,{a_2^2}\,
{a_4}\,b_4 - |d_{21}|^4}\;.
\end{eqnarray}

\subsection{Nonsymmetric case}\label{sec:LTNSymc}

We now turn to the case of a nonsymmetric critical end point, assuming
that $d_{11} > 0$. Owing to the implied explicit breaking of
the $\phi\to -\phi$ symmetry, the coexistence
surface $\boldsymbol{\rho}$ should neither be 
located in the $h=0$ plane nor be parallel to it.
Hence we must consider nonvanishing values of $h$ from
the outset.
Instead of (\ref{symm.GLEphi}) and (\ref{symm.GLEpsi}),
the classical equations of state (\ref{cleqstgen}) now read
\begin{eqnarray}
\label{nsym.GLEphi}
\left(a_2+d_{21}\, \psi\right) \phi +d_{11}\,\psi +
a_4\, \phi^3 &=& h\;,\\[0.5em]
\label{nsym.GLEpsi}
b_2\,\psi + b_4\,\psi^3 +d_{11}\,\phi +\frac{d_{21}}{2}\,\phi^2&=& g\;.
\end{eqnarray}

Using the solutions $\eta_{0,\pm}$ introduced in
 (\ref{eta0}), (\ref{eta0exp}), and (\ref{etapmexp}),
the former can be solved for $\phi$ to obtain
\begin{equation}\label{phins}
\phi=\eta_{0,\pm}{\left(a_2+d_{21}\,\psi,a_4,h-d_{11}\,\psi\right)}\;,
\end{equation}
where the subscript $0$ is to be taken whenever
$a_2+d_{21}\,\psi >0$, while the choices $\pm$ apply
to the cases $a_2+d_{21}\,\psi<0$ with
$h-d_{11}\,\psi\gtrless 0$, respectively. Likewise, 
solving the other equation of state, (\ref{nsym.GLEphi}), for
$\psi$ gives
\begin{equation}\label{psins}
\psi=\eta_{0,\pm}{\left(b_2,b_4,g-d_{11}\,
\phi-\frac{d_{21}}{2}\,\phi^2\right)}\;.
\end{equation}
Since we shall continue to take $b_2<0$, the subscripts $\pm$
are the appropriate choices, depending on whether
\begin{equation}
\check{g}(\phi)\equiv g-d_{11}\,\phi-(d_{21}/2)\,\phi^2
\end{equation}
is positive or negative.

From the above equations one can read off that 
the solutions $\phi_{\mathrm{min}}(\psi)$ given in ({\ref{phimin})
now apply on the hyperplane $h=d_{11}\,\psi$.
If $a_2+d_{21}\,\psi>0$, then any thermodynamically stable state
on this hyperplane must have $\phi=0$.
Furthermore, the liquid-gas critical point is seen to
remain at $b=g=h=0$.

Since $\phi_\lambda$, the equilibrium value
$\phi$ takes on the $\lambda$-line, generically
does not vanish, the fields $\phi$ and $\psi$
do no longer decouple there. To see this, let us
compute the
Hessian of ${\mathcal V}$,\begin{equation}
\boldsymbol{\mathcal V}^{(2)}(\phi,\psi)
\equiv
\left(\begin{array}{cc}{\mathcal V}_{\phi\phi}&
{\mathcal V}_{\phi\psi}\\ {\mathcal V}_{\psi\phi}&
{\mathcal V}_{\psi\psi}\end{array}\right)=
\left(\begin{array}{c@{\;\;}c}
\frac{\partial^2{\mathcal V}}{\partial\phi\,\partial\phi}&
\frac{\partial^2{\mathcal V}}{\partial\phi\,\partial\psi}\\[0.5em]
\frac{\partial^2{\mathcal V}}{\partial\phi\,\partial\psi}&
\frac{\partial^2{\mathcal V}}{\partial\psi\,\partial\psi}
\end{array}\right)\;,
\end{equation}
at a solution $\phi=\phi_{\mathrm{cl}},\,\psi=\psi_{\mathrm{cl}}$
of the equations of state (\ref{nsym.GLEphi}) and (\ref{nsym.GLEpsi})
(``classical solution''). This quantity
is nothing but the Fourier transform
\begin{equation}
\tilde{\boldsymbol{\Gamma}}^{(2)}{\left(
\boldsymbol{q}\right)}=
\left(\int\!d^dx_{12}\,
\Gamma_{\mu\nu}{\left(\boldsymbol{x}_1,\boldsymbol{x}_2;
\right)}\,
{\mathrm e}^{{\mathrm i}\boldsymbol{q}\cdot\boldsymbol{x}_{12}}
\right)
\end{equation}
at momentum $\boldsymbol{q}=\boldsymbol{0}$ of
the familiar Ornstein-Zernicke expression for the
two-point vertex function
\begin{equation}
\Gamma_{\mu\nu}{\left(
\boldsymbol{x}_1,\boldsymbol{x}_2\right)}\equiv
\frac{\delta^2\Gamma[\phi{=}\phi_{\mathrm{cl}},\psi{=}\psi_{\mathrm{cl}}]}%
{\delta\mu(\boldsymbol{x}_1)\,\delta\nu(\boldsymbol{x}_2)}
\;.
\end{equation}
Here the indices $\mu,\,\nu$ take the values $\phi,\,\psi$. Further, $\boldsymbol{x}_{12}$
means the deplacement vector $\boldsymbol{x}_1-\boldsymbol{x}_2$,
and $\Gamma[\phi,\psi]={\mathcal H}[\phi,\psi]$
in the Landau approximation used here.

As is borne out by the result
\begin{equation}\label{V2}
\boldsymbol{\mathcal V}^{(2)}(\phi_{\mathrm{cl}},\psi_{\mathrm{cl}})=
{\left({\begin{array}{c@{\;\;}c}
a_2 +d_{21}\,\psi_{\mathrm{cl}} + 3\,a_4\,\phi_{\mathrm{cl}}^2&
d_{11} + d_{21}\,\phi_{\mathrm{cl}}\\[0.5em]
d_{11} + d_{21}\,\phi_{\mathrm{cl}}&
b_2 + 3\,b_4\,\psi_{\mathrm{cl}}^2
\end{array}}\right)},
\end{equation}
the Hessian now is generically {\it nondiagonal\/},
even in the $\phi=0$ plane (where it is
diagonal in the symmetric case).

A principal axis transformation
with the orthogonal transformation matrix
\begin{eqnarray}
\boldsymbol{U}&=&\left(\begin{array}{c@{\;\;}c}
\cos{\vartheta}&\sin{\vartheta}\\[0.5em]
-\sin{\vartheta}
&\cos{\vartheta}
\end{array}\right)
\end{eqnarray}
yields the diagonal matrix
\begin{equation}\label{diag}
\mathrm{diag}\left(\lambda_1,\lambda_2\right)=
\boldsymbol{U}^{\mathrm{T}}\cdot
\boldsymbol{\mathcal V}^{(2)}
(\phi_{\mathrm{cl}},\psi_{\mathrm{cl}})\cdot
\boldsymbol{U}
\end{equation}
with the eigenvalues
\begin{eqnarray}
\lambda_{1\atop 2}&=&\frac{{\mathcal V}_{\phi\phi}+{\mathcal V}_{\psi\psi}}{2}
\mp\frac{1}{2}\,\sqrt{\left(%
{\mathcal V}_{\psi\psi}-{\mathcal V}_{\phi\phi}
\right)^2+4\,{{\mathcal V}_{\phi\psi}}^2}\;,
\end{eqnarray}
where the angle $\vartheta$ is given by
\begin{equation}\label{tanvartheta}
\tan{2\vartheta}=\frac{2\,{\mathcal V}_{\phi\psi}}{{\mathcal V}_{\psi\psi}-
{\mathcal V}_{\phi\phi}}\;.
\end{equation}
We write the associated eigendensities as
\begin{equation}\label{eigenfields}
\left(\begin{array}{c}\varphi_1\\%[0.5em]
\varphi_2\end{array}\right)
=\boldsymbol{U}^{\mathrm{T}}\cdot
\left(\begin{array}{c}\phi-\phi_{\mathrm{cl}}\\%[0.5em]
\psi-\psi_{\mathrm{cl}}\end{array}\right)\,.
\end{equation}

On the $\lambda$-line, the Hessian
$\boldsymbol{\mathcal V}^{(2)}(\phi_\lambda,\psi_\lambda)$
must have one vanishing eigenvalue, $\lambda_1$, and
a positive one, $\lambda_2$:
\begin{equation}\label{condlambda1}
\lambda_1(\phi_\lambda,\psi_\lambda)=0\;,\quad 
\lambda_2(\phi_\lambda,\psi_\lambda)>0\;.
\end{equation}
In order that ${\mathcal{V}}(\phi,\psi)>{\mathcal V}(\phi_\lambda,\psi_\lambda)$ for small deviations $\varphi_1$
with $\varphi_2=0$, the third derivative of ${\mathcal{V}}$
along the eigendirection $\varphi_1$,
\begin{eqnarray}\label{v30}
\left.{\partial^3{\mathcal{V}}
\over \partial\varphi_1^3}
\right|_{\mathrm{cl}}
&=&6\,a_4\,\phi_{\mathrm{cl}}\,\cos^3{\vartheta}
-3\,d_{21}\,\cos^2{\vartheta}\,\sin{\vartheta}
\nonumber\\&&\mbox{}
-6\,b_4\,\psi_{\mathrm{cl}}\,\sin^3{\vartheta}\;,
\end{eqnarray}
must vanish on the $\lambda$-line:
\begin{equation}\label{varphi13coeff}
\left.{\partial^3{\mathcal V}\over \partial\varphi_1^3}
\right|_\lambda
=0\;,
\end{equation}
while the corresponding fourth derivative
\begin{eqnarray}
\left.
{\partial^4{\mathcal V}\over \partial\varphi_1^4}
\right|_{\mathrm{cl}}
&=&6\,a_4\,\cos^4{\vartheta}
+6\,b_4\,\sin^4{\vartheta}
\end{eqnarray}
has to be positive there. The latter is
guaranteed by our assumptions that
$a_4>0$ and $b_4>0$.

Finally, the analog of
the requirement (\ref{a4eff}) that the effective
${\phi}^4$ coupling constant be positive becomes
\begin{equation}
\frac{1}{6}\left[\left.{d^4\over d\varphi_1^4}
\right|_{\varphi_2^{\mathrm{min}}(\varphi_1)}
{\mathcal V}\right]_{\mathrm{cl}}\equiv
\check{a}_4
(\phi_{\mathrm{cl}},\psi_{\mathrm{cl}})>0\;,
\end{equation}
where the derivative on the left-hand side is along a path
$\varphi_1\mapsto\varphi^{\mathrm{min}}_2(\varphi_1)$
through $(\phi_{\mathrm{cl}},\psi_{\mathrm{cl}})$
on which $\varphi_2$ takes that value
$\varphi_2^{\mathrm{min}}(\varphi_1)$
which minimizes ${\mathcal{V}}$ for given $\varphi_1$.
Exploiting the fact that
\begin{equation}
\frac{\partial{\mathcal{V}}}{\partial\varphi_1}\,d\varphi_1=
-\frac{\partial{\mathcal{V}}}{\partial\varphi_2}\,d\varphi_2
\end{equation}
on this path, one is led to the result
\begin{eqnarray}\label{varphi14iac}
\check{a}_4(\phi_{\mathrm{cl}},\psi_{\mathrm{cl}})
&=&\left[\frac{1}{6}\,
{\partial^4{\mathcal V}\over \partial\varphi_1^4}
-\frac{1}{2\,\lambda_2}\,\left({\partial^3{\mathcal{V}}
\over \partial\varphi_1^3}\right)^2
\right]_{\mathrm{cl}}\,,
\end{eqnarray}
which is compatible with, and generalizes,
(\ref{a4eff}).

Let $a_{2\lambda}$ and $h_\lambda$ be the values
$a_{2}$ and $h$ take as a function of $g$ and the other
variables on the $\lambda$-line. These values are
fixed by the first one of
the conditions (\ref{condlambda1}),
$\lambda_1=0$, and (\ref{varphi13coeff}).
To determine the coexistence surfaces $\boldsymbol{\rho}$
and $\boldsymbol{\sigma}$, one must again exploit the equality of
the grand potentials of the corresponding coexisting phases
[cf.\ (\ref{grandpot})].

Since $\phi_\lambda$ now does no longer vanish,
these conditions are not easy to handle analytically.
In general, one must recourse to numerical methods.
Fig.~\ref{fig:pdns} shows an example of a phase diagram
obtained in this fashion for the choice of interaction constants
given in the caption. As expected, the coexistence surface
$\boldsymbol{\rho\/}$, and hence the critical end point, are
displaced from the $h=0$ plane. Further, $\boldsymbol{\rho\/}$
gets curved, and the reflection symmetry of the phase diagram
with respect to $\boldsymbol{\rho\/}$
we had in the symmetric case is lost.

\begin{figure}[htb]
\resizebox{\columnwidth}{!}{%
  \includegraphics{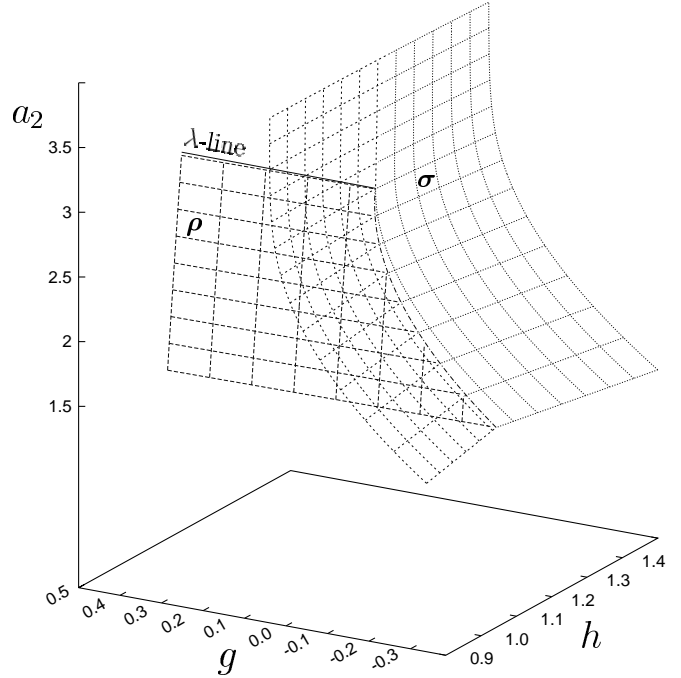}
}
\caption{Phase diagram of the nonsymmetric model
in the Landau approximation. The results, obtained
by numerical (and analytical, see text) solutions of
the equations of state, correspond to the following
choice of parameter values:
$b_2=-6$, $a_4=b_4=1/6$, $d_{21}=-1/2$, $d_{11}=0.2$.
Only the neighborhood
of the critical end point is shown.}
\label{fig:pdns}      
\end{figure}

These features can also be verified
by analytic calculations. To this end, we will restrict
ourselves to the vicinity
of the critical end point and treat the
symmetry-breaking term $\propto d_{11}$
in a perturbative fashion. In order to determine the
$\lambda$-line, we take $b_2<0$ and $g\ge g_{\mathrm{e}}$,
with $g_{\mathrm{e}}=g_{\mathrm{e}}(d_{11})$,
the (as yet unknown) value of $g$ at the critical end point. We write
\begin{equation}\label{philgexp}
\phi_\lambda(g)=\phi^{(1)}_\lambda(g)\,{d_{11}}+{O}(d_{11}^2)\;,
\end{equation}
\begin{equation}\label{psilgexp}
\psi_\lambda(g)={\psi_+}(g)+\psi^{(2)}_\lambda(g)\;{d_{11}^2}+O(d_{11}^3)\;,
\end{equation}
and
\begin{equation}\label{a2lgexp}
a_{2\lambda}(g)=|d_{21}|\,\psi_+(g)+{O}(d_{11}^2)\;,
\end{equation}
where $\psi_+(g)$ was defined in (\ref{psi+g}).
The values of these quantities for $d_{11}=0$ are known from
our analysis of the symmetric critical end point in the previous
subsection. That the term linear in $d_{11}$ of $\psi_\lambda(g)$
vanishes follows from the fact that 
$\phi_\lambda$ and $\phi_\lambda$ are coupled via terms of order
$d_{11}\,\phi_\lambda\sim\phi_\lambda^2\sim d_{11}^2$.
Upon substituting these expansions into
 the condition (\ref{varphi13coeff}) and the
expression (\ref{tanvartheta}) for $\tan\vartheta_\lambda$,
we can solve for $\phi^{(1)}_\lambda(g)$, obtaining
\begin{eqnarray}
\phi^{(1)}_\lambda(g)&=&\frac{-|d_{21}|}{2\,a_4\,{\left[
3\,b_4\,{\psi_+^2}(g)-b_2\right]}-d_{21}^2}
\nonumber\\&=&
\frac{-|d_{21}|\,{\psi_+}'(g)}{2\,{\check{a}_4{\left[\psi_+(g)\right]}}}
\;,
\end{eqnarray}
where $\check{a}_4{\left[\psi_+(g)\right]}$ and ${\psi_+}'(g)$
are given by (\ref{a4eff}) and (\ref{psi+prime}), respectively.

The above expansions can now be inserted into the equation of state
(\ref{nsym.GLEphi}) to determine $h_\lambda$. This gives
\begin{equation}\label{hlpt}
h_\lambda(g)=d_{11}\,\psi_+(g)+O(d_{11}^3)\;.
\end{equation}
Utilizing the result (\ref{psins}) for $\psi_\lambda$,
one gets the second-order coefficient
\begin{equation}
\psi^{(2)}_\lambda(g)=-d_{11}^2\,{\psi_+}'(g)\,{\phi^{(1)}_\lambda}(g)
{\left[1-{|d_{21}|\over 2}\,{\phi^{(1)}_\lambda}(g)\right]}\;.
\end{equation}

In order to be able to determine $g_{\mathrm{e}}$
from the
coexistence condition
\begin{eqnarray}\label{coexcondge}
\lefteqn{{\mathcal V}{\left[\phi_\lambda,\psi_\lambda;
a_{2\lambda}(g_{\mathrm{e}}),h_\lambda(g_{\mathrm{e}}),
g_{\mathrm{e}}\right]}}&&\nonumber\\
&=&
{\mathcal V}{\left[\phi_\alpha,\psi_\alpha;
a_{2\lambda}(g_{\mathrm{e}}),h_\lambda(g_{\mathrm{e}}),
g_{\mathrm{e}}\right]}\;,
\end{eqnarray}
we must also know $\psi_\alpha$ and $\phi_\alpha$
to the appropriate order in $d_{11}$.
The corresponding series expansions are analogous to
the those of $\psi_\lambda$ and $\phi_\lambda$ given in
(\ref{psilgexp}) and (\ref{philgexp}). The zeroth-order
term of $\psi_\alpha$ is $\psi_-(g)\equiv -\psi_+(-g)$
[cf.\ (\ref{+-sym})].
Hence we have
\begin{equation}
{\left[h-{d_{11}}\,\psi_\alpha\right]}_\lambda=
{d_{11}}\,{\left[\psi_+(g)+\psi_+(-g)\right]}+O(d_{11}^3)
\end{equation}
and
\begin{equation}
{\left[a_2+{d_{21}}\,\psi_\alpha\right]}_\lambda=
|{d_{21}}|\,{\left[\psi_+(g)+\psi_+(-g)\right]}+O(d_{11}^2)\;.
\end{equation}
To obtain $\phi_\alpha$ on the $\lambda$-line,
we insert these results into the expression 
(\ref{phins}) for $\eta_0$,
which we expand in powers
of its last argument, ${{\left[h-{d_{11}}\,\psi_\alpha\right]}_\lambda}$,
using (\ref{eta0exp}). 
This yields
\begin{equation}\label{phialphl1}
{\left.\phi_\alpha\right|}_\lambda=\frac{d_{11}}{|d_{21}|}+O(d_{11}^3)\;,
\end{equation}
which leads to
\begin{equation}
{\left.\psi_\alpha\right|}_\lambda=-\psi_+(-g)
-{d_{11}^2\over 2\,|d_{21}|}\,{\psi_+}'(-g)
+O(d_{11}^3)
\end{equation}
upon insertion into the expression (\ref{psins}) for $\eta_-$.

Using the above results, one can determine
$g_{\mathrm{e}}(d_{11})$ to second order in $d_{11}$
in a straightforward manner
from the condition  (\ref{coexcondge}). The result is
\begin{equation}\label{gept}
g_{\mathrm{e}}(d_{11})=\frac{d_{11}^2}{2\,|d_{21}|}+O(d_{11}^3)\;.
\end{equation}

As a check, let us set the interaction constants to the values
$b_2=-6$, $a_4=b_4=1/6$, $d_{21}=-1/2$, $d_{11}=0.2$
utilized in our numerical analysis (see Fig.~\ref{fig:pdns}).
Our perturbative analytical results (\ref{gept}), (\ref{hlpt}), and
(\ref{a2lgexp}) then predict the critical end point to be located at
\begin{eqnarray}
g_{\mathrm{e}}&\simeq& 0.04\;,\nonumber\\
h_{\mathrm{e}}&=&h_{\lambda}(g_{\mathrm{e}})\simeq 1.2\;,\nonumber\\
a_{2\mathrm{e}}&=&a_{2\lambda}(g_{\mathrm{e}})\simeq 3.0\;,
\end{eqnarray}
and to have the slopes
\begin{eqnarray}
{h_{\lambda}}'(g_{\mathrm{e}})&\simeq &d_{11}\,{\psi_+}'(g_{\mathrm{e}})
\simeq 0.017\;,
\nonumber\\
{a_{2\lambda}}'(g_{\mathrm{e}})&\simeq &|d_{21}|\,{\psi_+}'(g_{\mathrm{e}})
\simeq 0.04\;.
\end{eqnarray}
These numbers are in excellent agreement with the numerical results
shown in Fig.~\ref{fig:pdns}.

Next, let us calculate the equilibrium
densities $\phi_{\beta,\gamma}$ and $\psi_{\beta,\gamma}$ of the ordered
phases $\beta$ and $\gamma$
in a perturbative manner. To this end, we consider
a path in the space $\{(a_2,h,g)\}$ with $g=\mbox{const}$
that is asymptotically parallel to the coexistence
surface $\boldsymbol{\rho}$
and intersects the $\lambda$-line at a point ${\mathcal Q}=
(a_{2\lambda}(g),h_{\lambda}(g),g)$
close or equal to the critical end point.
Taking ${\mathcal Q}$ as expansion point in (\ref{eigenfields}), we express
the deviations $\delta \phi\equiv\phi-\phi_{\lambda}$ and $\delta \psi\equiv\psi-\psi_{\lambda}$ from the associated classical solution
$(\phi_{\mathrm{cl}},\psi_{\mathrm{cl}})=(\phi_{\lambda},\psi_{\lambda})$
in terms of the eigendensities $\varphi_1$ and $\varphi_2$ of ${\mathcal{Q}}$.

The shifts $\delta a_2\equiv a_2-a_{2\lambda}(g)$ and
$\delta h=h-h_\lambda(g)$ induce the following changes
of ${\mathcal{V}}(\phi,\psi)$:
Nonvanishing  terms
\begin{equation}
{\frac{{\delta a_2}}{2}}\,{\cos(\vartheta_\lambda})\,\varphi_1^2=
\frac{{\delta a_2}}{2}\,{\left[1+O(d_{11}^2)\right]}\varphi_1^2\;,
\end{equation}
\begin{equation}
\frac{1}{2}\,{\delta a_2}\,{\varphi_1}{\varphi_2}\,
\sin{2\vartheta_\lambda}\sim d_{11}\,\delta a_2\,{\varphi_1}{\varphi_2}\;,
\end{equation}
and\footnote{Note that $\delta h\sim d_{11}$. Its zeroth-order term
in $d_{11}$ vanishes
because the coexistence surface is given by $h=0$ in this (symmetric) case.}
\begin{equation}
{\varphi_2}\,{\left[{
\left(\phi_\lambda\,\delta a_2-{\delta h}\right)}\,
{\sin{\vartheta_{\lambda}}}\right]}\sim d_{11}^2\,\varphi_2
\end{equation}
are generated;
the coefficient $\lambda_2(\phi_\lambda,\psi_\lambda)$
of the ${1\over 2}\,{\varphi_2}^2$ term is changed by an amount $\delta a_2\,O(d_{11}^2)$.
The term linear in $\varphi_1$
is negligible
sufficiently close to ${\mathcal{Q}}$ since we required the
path to be asymptotically parallel to the coexistence surface $\boldsymbol{\rho}$.

Minimizing with respect to $\varphi_1$ and $\varphi_2$
yields the equilibrium values
\begin{eqnarray}
\varphi_{1{\beta\atop\gamma}}&=&\pm
\sqrt{\frac{\left|{\delta{a}_2}\right|}
{\check{a}_{4\lambda}}}\;%
\left[1+o{(\delta{a}_2)}\right]\;,\\
\varphi_{2{\beta\atop\gamma}}&=&\delta a_2\,O(d_{11}^2)
\;,
\end{eqnarray}
from which we get
\begin{eqnarray}
\phi_{{\beta\atop\gamma}}-\phi_\lambda&=&\pm
\sqrt{\frac{\left|{\delta{a}_2}\right|}
{\check{a}_{4\lambda}}}\;%
\left[1+o{(\delta{a}_2)}\right]
\end{eqnarray}
and
\begin{eqnarray}
\psi_{{\beta\atop\gamma}}-\psi_\lambda&\approx&{\psi_+}'(g)\,{\left[
|{d_{21}}|\,{\frac{|{\delta{a}_2}|}{\check{a}_{4\lambda}}}
\mp d_{11}\,\sqrt{\frac{|{\delta{a}_2}|}{\check{a}_{4\lambda}}}
\right]}.
\end{eqnarray}
Since $\delta{a_2}\sim (-t)$, the result displays the
expected $|t|^\beta$ singularity of (\ref{singdens}),
with the mean-field exponent $\beta= {1/2}$.

In order to determine the  wings of  $\alpha$-$\beta$ and $\alpha$-$\gamma$
coexistence of the first-order surface $\boldsymbol{\sigma}$,
we would have to substitute the classical equilibrium solutions $\phi_\wp$
and $\psi_\wp$, with $\wp=\alpha$, $\beta$, $\gamma$,
into ${\mathcal{V}}(\phi,\psi)$ to obtain the corresponding
grand potentials ${\mathcal{A}}_{\wp}(a_2,h,g)$ and then
solve the coexistence conditions
\begin{equation}\label{abcoexcond}
{\mathcal{A}}_{\alpha}(a_2,h,g)={\mathcal{A}}_{l}(a_2,h,g)\;,
\quad l=\beta,\gamma\,,
\end{equation}
e.g., for $g$ as a function of $a_2$ and $h$. On the triple line
conditions (\ref{abcoexcond}) hold for both $l=\alpha$ and $l=\beta$.

Let us consider small deviations from the critical end point that are
directed along the triple line.
We parametrize these through $\delta a_2\equiv
a_2-a_{2\mathrm{e}}$ and denote the value of the effective $\varphi_1^4$
interaction constant (\ref{varphi14iac}) at the critical end point
as $\check{a}_{4\mathrm{e}}$. Using the above results
and neglecting terms of order $d_{11}^2$, it is
not difficult to show that the triple-line value $g_\sigma$
behaves as
\begin{equation}\label{gtriple}
g_\sigma\approx-\left[1+O{\left(d_{11}^2\right)}\right]
\frac{\left(\delta{a_2}\right)^2}%
{8\,\check{a}_{4\mathrm{e}}\,\sqrt{{|b_2|/b_4}}}
\end{equation}
as $\delta a_2\sim t\to -0$. The second derivative of
this expression with respect to $t$, taken at $t=+0$, gives
us the expected jump singularity.
(The corresponding value of $g''_\sigma(t)$
at $t=+0$ is zero.)

\section{Beyond Landau theory}\label{sec:beyLT}

In the previous section we saw that 
 Landau theory yields for both the symmetric
and nonsymmetric versions of our model  phase diagrams with the correct
features. We also verified that the results of this approximation
are consistent with the
predicted singularities (\ref{singpb}) and (\ref{singdens}).
We now wish to extend the analysis beyond Landau theory.

It will be helpful to make a number of
remarks before we embark on details.
Our strategy will be to relate
the critical singularities on both the critical line \emph{and}
at the critical end point to those of the standard $\phi^4$ theory.
In other words, we will show that they can be described by
a Hamiltonian of the form (\ref{H1}) of ${\mathcal{H}}_1[\phi]$.
To understand how this works, it is useful to consider first the
theories described by the
Hamiltonians $\mathcal {H}_1[\phi]$ and ${\mathcal{H}}_2[\psi]$
in the absence of any coupling between $\phi$ and $\psi$, and
then discuss what happens when ${\mathcal{H}}_{12}[\phi,\psi]$
is turned on.

\subsection{The decoupled noncritical theory %
and redundant operators}\label{sec:RedOp}

Suppose that $B>0$, $b_4>0$, $g=\pm 0$, and $b_2<b_{2c}$, where
$b_{2c}$ is the critical value  below which the symmetry $\psi\to -\psi$ 
is spontaneously broken.
(At the level of Landau theory, $b_{2c}=0$, of course.)
Then ${\mathcal{H}}_2[\psi]$ describes a \emph{massive} field theory
whose statistical properties
on length scales large
compared to the corresponding correlation length $\xi_\psi$
may be characterized by a Gaussian probability distribution
with Hamiltonian
\begin{equation}\label{Gauss2}
\frac{w_2}{2}\,{\mathcal{H}}_\mathrm{G}[\delta\psi]=
\frac{w_2}{2}{\int_\Omega}\left[\psi - \psi^{(\pm)}\right]^2\;.
\end{equation}
Here $\psi^{(\pm)}$ are the mean values of $\psi$ in the pure
phases selected by the choices $g=\pm 0$, respectively, while
$\delta\psi$ denotes the fluctuating quantity inside the brackets
on the right-hand side of (\ref{Gauss2}). Hence the correlation
function becomes
\begin{equation}
\left\langle\delta\psi(\boldsymbol{x})\,\delta\psi(\boldsymbol{x}')\right\rangle
= w_2\,\delta (\boldsymbol{x}-\boldsymbol{x}')
\end{equation}
for $|\boldsymbol{x}-\boldsymbol{x}'|\gg \xi_\psi$. 

If we let $B\to 0$ in ${\mathcal{H}}_2[\psi]$, so that
$\xi_\psi\to 0$, and ignore the cubic and quartic terms in (\ref{H2pm}),
we obtain $w_2=2|b_2|$ and $\psi^{(\pm)}=\pm\sqrt{|b_2|/b_4}$.
Inclusion of these terms (e.g., by means of perturbation theory)
does not modify this large-scale form of the theory's correlation functions,
 but produces different values of
$w_2$ and $\psi^{(\pm)}$. In other words, the parameters of this
Gaussian (fixed-point) Hamiltonian are changed, albeit its form
remains the same. In RG theory such a change of coordinates
of a fixed-point Hamiltonian in parameter space is known
to be caused by \emph{redundant} operators \cite{Weg76}.

For the sake of clarity, let us briefly recall the concept of
redundant operators. Consider a field theory with a Hamiltonian
${\mathcal{H}}[\varphi]$, where $\varphi$ could be a single-component field,
such as $\phi$ or $\psi$, or a multi-component field $(\varphi^\alpha)$,
such as $(\phi,\psi)$. Suppose ${\mathcal{H}}^*[\varphi]$ is a fixed point
under RG transformations. We consider operators
of the form
\begin{equation}\label{op}
{\mathcal{O}}={\int_\Omega}\!d^dx\, {\mathcal{O}}(\boldsymbol{x})\;,\quad
{\mathcal {O}}(\boldsymbol{x})={O}[\varphi(\boldsymbol{x}),
\nabla\varphi(\boldsymbol{x})]\;,
\end{equation}
i.e., ${O}(\boldsymbol{x})$ is a local operator depending on $\varphi$
and its derivatives, but not explicitly on $\boldsymbol{x}$
(to ensure translational invariance.)
Such an operator is called redundant if a local functional
$\Upsilon(\boldsymbol{x},[\varphi])=
\Upsilon[\varphi(\boldsymbol{x}),\nabla\varphi(\boldsymbol{x})]$
exists such that it can be written as
\begin{equation}
{\mathcal{O}}_\mathrm{red}={\mathfrak{G}_{\mathrm{tra}}}[\Upsilon]\,
{\mathcal{H}^*}[\varphi]\;,
\end{equation}
where $\mathfrak{G}_{\mathrm{tra}}[\Upsilon]$ is the generator
of transformations of ${\mathcal{H}}$ induced by the change of variable
\begin{equation}\label{ftransf}
\varphi(\boldsymbol{x})=
\varphi'(\boldsymbol{x})+\Upsilon(\boldsymbol{x},[\varphi'])\;.
\end{equation}
That is,
\begin{equation}
\mathfrak{G}_{\mathrm{tra}}[\Upsilon]\,{\mathcal{H}}[\varphi]
\equiv{\int_\Omega} \!d^dx
\left[\Upsilon(\boldsymbol{x})
\frac{\delta{\mathcal{H}}[\varphi]}{\delta\varphi(\boldsymbol{x})}
-\frac{\delta\Upsilon(\boldsymbol{x})}{\delta\varphi(\boldsymbol{x})}\right],
\end{equation}
where the second term in brackets results from the Jacobian of the transformation
(\ref{ftransf}).

Now, consider the noncritical Gaussian Hamiltonian (\ref{HG})
or (\ref{Gauss2}). This is a fixed point under RG transformations.
Wegner (see Sec.\ III.G.2 of Ref.\ \cite{Weg76}) has shown that
\emph{any operator ${\mathcal{O}}$
of the form (\ref{op})} --- i.e., the integral of
any local operator --- 
\emph{at this Gaussian fixed point can be expressed as a sum
of a redundant operator and a constant}:
\begin{equation}\label{redop}
{\mathcal{O}}={\mathcal{O}}_{\mathrm{red}}+ C_{\mathcal{O}}\,|\Omega|\;,
\end{equation}
where the constant $C_{\mathcal{O}}$ may be zero for some operators
(as it must for any operator that is odd in $\psi$).
The result (\ref{redop}) can be proven in a constructive
manner \cite{rem:redproof}. We content ourselves to
showing how (\ref{redop}) reads for the cubic, quartic, and
$(\nabla\psi)^2$ operators appearing in (\ref{H2pm}); one has
\begin{equation}
{\int_\Omega} \psi^3=\frac{1}{2}\,\mathfrak{G}_{\mathrm{tra}}[\psi^2+2\,\delta(\boldsymbol{0})]\,
{\mathcal{H}}_\mathrm{G}[\psi]\;,
\end{equation}
\begin{equation}
{\int_\Omega} \psi^4=\frac{1}{2}\,\mathfrak{G}_{\mathrm{tra}}[\psi^3+3\,\psi\,\delta(\boldsymbol{0})]\,
{\mathcal{H}}_{\mathrm{G}}[\psi]-3|\Omega|\,[\delta(\boldsymbol{0})]^2\;,
\end{equation}
and
\begin{equation}
{\int_\Omega} (\nabla\psi)^2=\frac{1}{2}\,\mathfrak{G}_{\mathrm{tra}}[-\nabla^2\psi]
\,{\mathcal{H}}_\mathrm{G}[\psi]
+|\Omega|\int_{\boldsymbol{q}}q^2\,.
\end{equation}
In a theory regularized by a cutoff $\Lambda$ the momentum integration
$\int_{\boldsymbol{q}}=\int d^dq/(2\pi)^d$ is restricted
by $|\boldsymbol{q}|\le \Lambda$; thus $\delta(\boldsymbol{0})=\int_{\boldsymbol{q}}1$ is not
infinite but equal to $K_d\,\Lambda^d$, with $K_d=2^{1-d}\pi^{-d/2}/\Gamma(d/2)$.

It is useful to recall the physical significance of the result (\ref{redop}):
The massive field theory described by ${\mathcal{H}}_2[\psi]$ does not
exhibit critical singularities; the associated bulk free energy density
is an analytic function of the thermodynamic fields.
Thus no scaling operators other than the trivial one $|\Omega|$ appear;
the associated scaling field is the previously mentioned special one
with RG eigenexponent $d$.

\subsection{The decoupled critical theory}\label{sec:decCT}

We next turn to the field theory described by the Hamiltonian
${\mathcal{H}}_1[\phi]$.
For given positive values of $A$ and $a_4$, this has a critical point
at $(a_2,h)=(a_{2c},0)$, where $a_{2c}=0$ in the Landau approximation.
The asymptotic behavior at this critical point can be
analyzed by means of well-known RG methods
\cite{ZJ96,BLZ76,Ami89}. In order to establish our notation,
it is necessary that we briefly recapitulate some of its ingredients.

Let us introduce the corresponding bulk free energy (per volume and $k_{\mathrm{B}}T$)
\begin{equation}\label{f1b}
f_{1,{\mathrm{b}}}(a_2,a_4,h)=-\lim_{\Omega\uparrow {\mathbb{R}\/}^d}
{\left\{\frac{1}{|{\Omega}|}
\ln {\int}{{\mathcal{D}}\phi}\,{\mathrm{e}}^{-{\mathcal{H}}_1[\phi]}\right\}}
\end{equation}
as well as the generating functionals ${\mathcal{Z}}_1[J,I]$
and
 ${\mathcal{G}}_1[J,I]$
of correlation functions and cumulants, respectively,
via
\begin{equation}\label{gfuncdef}
{\mathcal{Z}}_1[J,I]=\mathrm{e}^{{\mathcal{G}}_1[J,I]}=
\left\langle
\mathrm{e}^{(J,\phi)+{1\over 2}\left(I,\phi^2\right)}
\right\rangle_{{\mathcal{H}}_1} \;,
\end{equation}
where the subscript ${\mathcal{H}}_1$ serves to indicate that the average
\begin{equation}\label{av1}
\left\langle\,.\,\right\rangle_{{\mathcal{H}}_1}\equiv
{{\int}{\mathcal{D}}{\phi}\,.\;\mathrm{e}^{-{\mathcal{H}}_1[\phi]}
\over 
{\int}{\mathcal{D}}{\phi}\,\mathrm{e}^{-{\mathcal{H}}_1[\phi]}
}
\end{equation}
refers to the decoupled system with  Hamiltonian ${\mathcal{H}}_1[\phi]$
(rather than to the full one with Hamiltonian ${\mathcal{H}}[\phi,\psi]$),
and
\begin{equation}\label{shthd}
(I,\phi^2)\equiv {\int_\Omega}\!d^dx\,I(\boldsymbol{x})\,\phi^2(\boldsymbol{x})
\end{equation}
is a convenient shorthand.

To absorb the ultraviolet (uv) divergences of the
theory in $d=4-\epsilon\le 4 $ dimensions, we  use reparametrizations
of the form
\begin{eqnarray}\label{DefZfactors}
\phi&=&\left[Z_\phi(u)\right]^{1/2}\,\phi_\mathrm{ren}\;,\\
a_{2}&=&\kappa^2\,Z_{\tau}(u)\,\tau+a_{2,c}\;,\\
a_4&=&\kappa^\epsilon\,Z_{u}(u)\,u\;,\\
h&=&\kappa^{(d+2)/2}\,{{\left[Z_\phi(u)\right]}^{-1/2}}\,{h_\mathrm{ren}}\;.\label{hren}
\end{eqnarray}
Here $\kappa$ is an arbitrary momentum scale. The 
renormalization factors are understood to be fixed my means of a
$\tau$  (`mass') independent renormalization scheme; for concreteness, we
shall assume that
dimensional regularization is employed and that they are
determined by minimal subtraction of poles in $\epsilon$.
In a cutoff-regularized theory, the shift $a_{2,c}$ of the
critical point from its zero-loop value $0$ diverges as $\Lambda^2$ as
$\Lambda\to\infty$.

In addition to the reparametrizations (\ref{DefZfactors})--(\ref{hren}), additive
counterterms are required. The bulk free energy $f_{1,{\mathrm{b}}}$
and its first and second
derivatives with respect to $a_2$ have primitive divergences
$\sim\Lambda^4$, $\sim\Lambda^2$ and $\sim\ln\Lambda$ at $d=4$,
respectively. To cancel these, we make subtractions at the normalization
point (NP) $a_2=a_2^{\mathrm{NP}}$, $h=0$, with
\begin{equation}
a_2^\mathrm{NP}=a_{2,c}+\kappa^2\,Z_\tau\,\tau^\mathrm{NP}\;,\quad
\tau^\mathrm{NP}=1\;.
\end{equation}
Accordingly we introduce the renormalized bulk free energy
\begin{eqnarray}\label{fb1ren}
f_{1,{\mathrm{b}}}^{\mathrm{ren}}(\tau,u,{h_\mathrm{ren}})
&=&f_{1,{\mathrm{b}}}(a_2,a_4,h)-
f_{1,{\mathrm{b}}}^{\mathrm{NP}}(a_4)
\nonumber\\&&\mbox{}
-{\left(a_2-a_2^\mathrm{NP}\right)}\,\varepsilon^{\mathrm{NP}}(a_4)
\nonumber\\&&\mbox{}
+ \frac{1}{2}\,{\left(a_2-a_2^\mathrm{NP}\right)}^2
\,C^\mathrm{NP}(a_4)\;,
\end{eqnarray}
where the renormalization functions $\varepsilon^{\mathrm{NP}}$ and $C^{\mathrm{NP}}$ are fixed through the normalization
conditions
\begin{equation}
f_{1,{\mathrm{b}}}^{\mathrm{ren}}(1,u,0)=0\;,
\end{equation}
\begin{equation}
{\partial f_{1,{\mathrm{b}}}^{\mathrm{ren}}\over\partial\tau}
(1,u,0)=0\;,
\end{equation}
and
\begin{equation}
{\partial^2 f_{1,{\mathrm{b}}}^{\mathrm{ren}}\over\partial\tau^2}
(1,u,0)=0\;.
\end{equation}
For the generating functional of renormalized cumulants
we have
(cf.\ Sec.\ 12.1.1 of  Ref.\ \cite{ZJ96})
\begin{eqnarray}\label{genfctG1}
{{\mathcal{G}}_{1}^{\mathrm{ren}}}[J,I]&=&
{{\mathcal{G}}_1}{\left[Z_\phi^{-1/2}J,Z_{\tau}\,I\right]}
-{\int_\Omega}{\bigg\{}{\varepsilon^\mathrm{NP}}Z_\tau \,I
\nonumber\\&&\mbox{}
- \frac{1}{2}\,C^\mathrm{NP}\,{\big[Z_\tau\,I-
\left(a_2-a_2^\mathrm{NP}\right)€\big]}^2{\bigg\}}.\;\;
\end{eqnarray}
Thus the renormalized cumulants
\begin{equation}
G_{1,\mathrm{ren}}^{(N,M)}=\left.\frac{\delta^{N+M}%
{\mathcal{G}}_{1}^{\mathrm{ren}}[J,I]}{{\delta J}(\boldsymbol{x}_1)\ldots{\delta J}(\boldsymbol{x}_N)\,{\delta I}(\boldsymbol{X}_1)\ldots{\delta I}(\boldsymbol{X}_M)}\,
%\left[\prod_{j=1}^M\frac{\delta}{\delta I(\boldsymbol{X}_j)}\right]
\right|_{J=0\atop I=0}
\end{equation}
are related to their analogously defined bare counterparts $G_{1}^{(N,M)}$
via
\begin{eqnarray}\label{rencum}
G_{1,\mathrm{ren}}^{(N,M)}&=&Z_\phi^{-N/2}\,Z_\tau^{M}\,
\Big\{G_{1}^{(N,M)}\nonumber\\&&\mbox{}\quad
-\delta_{M,1}^{N,0}\left[
\varepsilon^\mathrm{NP}-C^\mathrm{NP}\,\kappa^2\,Z_\tau
\left(\tau-1\right)
\right]\nonumber\\&&\mbox{}\qquad\quad
+\delta_{M,2}^{N,0}\,
C^\mathrm{NP}\,\delta({\boldsymbol{X}}_{12})
\Big\}\,,
\end{eqnarray}
where ${\boldsymbol{X}}_{12}=\boldsymbol{X}_1-\boldsymbol{X}_2$.

Upon varying $\kappa$, we can derive RG equations.
Let us introduce the beta function
\begin{equation}
\beta_u(u)=\left.\kappa\partial_\kappa\right|_0u\;,
\end{equation}
the exponent functions
\begin{equation}
\eta_\phi(u)=\left.\kappa\partial_\kappa\right|_0\,\ln Z_\phi\;,
\end{equation}
\begin{equation}
\eta_\tau(u)=\left.\kappa\partial_\kappa\right|_0\,\ln Z_\tau\;,
\end{equation}
and the differential operator
\begin{equation}\label{Dkappa}
{\mathcal{D}}_\kappa\equiv \kappa\partial_\kappa+\beta_u\partial_u-(2+\eta_\tau)\tau\partial_\tau
-\frac{d{+}2{-}\eta_\phi}{2}\,h_\mathrm{ren}\partial_{h_{\mathrm{ren}}}\,,
\end{equation}
where $\partial_\kappa|_0$ means a derivative at fixed
values of the bare parameters $a_2$, $a_4$, and $h$.
Then the RG equations can be written as
\begin{equation}\label{RGEfb}
{\mathcal{D}_\kappa}\,f_{b,1}^{\mathrm{ren}}=
-\frac{2+\eta_\tau}{2}\,{\left(\tau-1\right)}^2
\left(\partial_\tau^3 f_{1,\mathrm{b}}^{\mathrm{ren}}\right)^{\mathrm{NP}}
\end{equation}
and
\begin{eqnarray}\label{RGE}
\lefteqn{\left[{\mathcal{D}}_\kappa
+{N\over 2}\,\eta_\phi
-M\,\eta_\tau\right]G^{(N,M)}_{1,\mathrm{ren}}
}\nonumber\\
&=&-{2+\eta_\tau\over\kappa^{2M}}\,
\left(\partial_\tau^3 f_{1,\mathrm{b}}^{\mathrm{ren}}\right)^{\mathrm{NP}}
{\left[{\delta_{M,1}^{N,0}}\,
{\left(\tau-1\right)}
-\delta_{M,2}^{N,0}
\right]}.
\nonumber\\
\end{eqnarray}

The RG equations (\ref{RGEfb}) and (\ref{RGE})
can be exploited in a standard fashion to
derive the familiar scaling forms of the (singular part of the) free energy and
of the cumulants $G^{(N,M)}_{1,\mathrm{ren}}$.
Details can be found, for instance,
in Refs.\ \cite{ZJ96,BLZ76,Ami89} or \cite{Die86a}
and will not be repeated here.
In the cases  of  the free energy $f_{1,\mathrm{b}}^{\mathrm{ren}}$,
the energy density $\varepsilon^\mathrm{ren}=G^{(0,1)}_{1,{\mathrm{ren}}}$,
and $G^{(0,2)}_{1,{\mathrm{ren}}}$ (specific heat)
the RG equations are inhomogeneous. Explicit
solutions to such equations in terms of RG trajectory integrals
are given in Ref.\ \cite{DD81a} (and elsewhere).
Specifically for the renormalized free energy
$f_{1,{\mathrm{b}}}^{\mathrm{ren}}$, the solution
reads
\begin{eqnarray}
f_{1,{\mathrm{b}}}^{\mathrm{ren}}(\tau,h_{\mathrm{ren}},u,\kappa)
&=&(\kappa\ell)^d\,f_{1,{\mathrm{b}}}^{\mathrm{ren}}
[\bar{\tau}(\ell),\bar{h}(\ell),\bar{u}(\ell),1]
\nonumber\\&&\mbox{}
+{\int_1^\ell}\frac{d\ell'}{\ell'}\,\bar{I}(\ell')\,.\;\;
\end{eqnarray}
with
\begin{eqnarray}
\lefteqn{%
\bar{I}(\ell)
=-\frac{2+\eta_\tau[\bar{u}(\ell)]}{2}\,{\left[\bar{\tau}(\ell)-1\right]}^2
\,{\left(\partial_\tau^3 f_{1,b}^{\mathrm{ren}}\right)}^{\mathrm{NP}}(\bar{u},\kappa\ell)}&&\nonumber\\
&=&-\frac{2+\eta_\tau[\bar{u}(\ell)]}{2}\,
{\kappa^d\,\ell^{3/\nu}\,{\left[\bar{\tau}(\ell)-1\right]}^2\over {\left[E_\tau(\bar{u}(\ell),u)\right]}^3}\,
{\left(\partial_\tau^3 f_{1,{\mathrm{b}}}^{\mathrm{ren}}\right)^{\mathrm{NP}}}(u,1)
\;,\nonumber\\
\end{eqnarray}
where $\bar{u}(\ell)$, $\bar{\tau}(\ell)$, and $\bar{h}(\ell)$ are standard
running variables [cf.\ Eqs.~(3.79a), (3.79.b), and (3.79d) of Ref.\ \cite{Die86a}].
Utilizing the notational conventions of this reference, we have
\begin{equation}
\bar{\tau}(\ell)=\ell^{-1/\nu}\,
E_\tau[\bar{u}(\ell),u]\,\tau \mathop{\approx}\limits_{\ell\to 0} 
\ell^{-1/\nu}\,E_\tau^*(u)\,\tau 
\end{equation}
and
\begin{equation}
\bar{h}(\ell)=\ell^{-\Delta/\nu}\,
E_h[\bar{u}(\ell),u]\,h\mathop{\approx}\limits_{\ell\to 0}
\ell^{-\Delta/\nu}\,
E_h^*(u)\,h_{\mathrm{ren}}\;.
\end{equation}
Here
\begin{equation}
E_{\tau}[\bar{u},u]=\exp\left\{
\int_u^{\bar{u}(\ell)}\!{dx}\,
{\eta_\tau^*-\eta_\tau(x)\over \beta_u(x)}
\right\}
\end{equation}
and
\begin{equation}
E_{h}[\bar{u},u]=\exp\left\{
\int_u^{\bar{u}(\ell)}\!{dx}\,
{\eta_\phi(x)-\eta\over \beta_u(x)}
\right\}\;.
\end{equation}
An asterisk is used to  mark values at the infrared-stable
fixed point $u^*$ of the beta function $\beta_u$. Further,
$E_{\tau,h}^*(u)$ means $E_{\tau,h}(u^*,u)$, and the exponents
are given by $\nu=(2{+}\eta_\tau^*)^{-1}$ and
$\Delta/\nu=(d{+}2{-}\eta_\phi^*)/2$, respectively.

The consequences of these equations for the
free energy $f^{\mathrm{ren}}_{1,\mathrm{b}}$
can be cast in the usual form: It is a sum of
a regular and a singular part,
\begin{equation}\label{fregsing}
f^{\mathrm{ren}}_{1,\mathrm{b}}
=f^{\mathrm{reg}}_{1,\mathrm{b}}
+f^{\mathrm{sing}}_{1,\mathrm{b}}\;.
\end{equation}
The latter, a solution of the homogeneous RG equation
of $f^{\mathrm{ren}}_{1,\mathrm{b}}$,
near criticality takes the scaling form
\begin{equation}\label{fsing}
f^{\mathrm{sing}}_{1,\mathrm{b}}(\tau,h_{\mathrm{ren}},u)\approx |\hat{\tau}|^{2-\alpha}\,Y_\pm(\hat{h}\,|\hat{\tau}|^{-\Delta})\;.
\end{equation}
Here $\hat{\tau}\approx E_\tau^*(u)\,\tau$ and
$\hat{h}\approx E_h^*(u)\,h_{\mathrm{ren}}$ are the two relevant
scaling fields. The scaling function $Y_\pm(y)$ is universal
(and hence independent of $u$); its two branches $\pm$ satisfy matching
conditions as $y\to\pm\infty$ (see, e.g., Ref.\ \cite{FU90b}).

\subsection{The coupled theory}\label{sec:CoupTh}
We are now ready to discuss the coupled theory, with
${\mathcal{H}}_{12}\ne 0$. We begin with the symmetric
case, setting $h=d_{11}=0$ and $A=1$. To simplify the
subsequent analysis,
we also set  the interactions constants $e_{21}$ and $f_{21}$
to zero, but will comment on their effects later [see the paragraph following
Eq.~(\ref{gradpsi2rw})].

\subsubsection{Symmetric case}{\label{sec:CoupThsym}

As reference densities about which we expand we take the
classical solutions $\phi_{\mathrm{cl}}=0$ and
$\psi_{\mathrm{cl}}$
at a point
$(a_2,h,g)=(a_{2\lambda}^{\mathrm{cl}}(g_{\mathrm{cl}}),0,g_{\mathrm{ref}})$ with
$g_{\mathrm{ref}}\ge g_e$ on the Landau-theory $\lambda$-line.
Writing
\begin{eqnarray}\label{defcheckpsi}
\psi&=&\psi_{\mathrm{cl}}+\check{\psi}\;,
\end{eqnarray}
we have
\begin{equation}\label{Hrewr}
{\mathcal{H}}[\phi,\psi]=
{\mathcal{H}}_2[\psi_{\mathrm{cl}}]+
\check{\mathcal{H}}[{\phi},\check{\psi}]+
{\mathcal{H}}'[{\phi},\check{\psi}]
\end{equation}
with
\begin{eqnarray}\label{Hcheck}
\check{\mathcal{H}}[{\phi},\check{\psi}]&=&
{\int_\Omega}{\left[
\frac{1}{2}\left(\nabla\phi\right)^2+{\check{a}_2\over 2}\,\phi^2
+{{a}_4\over 4}\,\phi^4
\right.}\nonumber\\&&{\left.\qquad\mbox{}
+{\check{b}_2\over 2}\,\check{\psi}^2
+\frac{d_{21}}{2}\,{\phi^2}\,\check{\psi}
\right]}
\end{eqnarray}
and
\begin{equation}\label{Hprime}
{\mathcal{H}}'[{\phi},\check{\psi};\psi_{\mathrm{cl}}]=
{{\int_\Omega}}{\left[
\frac{B}{2}\left(\nabla\check{\psi}\right)^2+{\check{b}_3
({\psi}_{\mathrm{cl}})\over 3}\,\check{\psi}^3
+{{b}_4
\over 4}\,\check{\psi}^4
\right]}\;,
\end{equation}
where
\begin{eqnarray}
\check{a}_2&=&a_2+{d_{21}}\,{\psi_{\mathrm{cl}}}\;,\label{acheck}\\
\check{b}_2&=&b_2+3{b_4}\,\psi_{\mathrm{cl}}^2\;,\\
\check{b}_3&=&3{b_4}\,{\psi_{\mathrm{cl}}}\;.
\end{eqnarray}

Suppose first that ${\mathcal H}'$ is neglected. Since $\check{\mathcal H}$
is quadratic in $\check{\psi}$,
having the same form as the Hamiltonian of the stochastic dynamic
models C, D, and E \cite{HHM72},  the field $\check{\psi}$ can then
be integrated out exactly to obtain an effective $\phi^4$ model.
Let us define quite generally, both  for ${\mathcal{H}}'=0$
as well as for ${\mathcal{H}}'\ne 0$, an effective Hamiltonian through
\begin{equation}\label{Hphieffdef}
{\mathrm{e}}^{-{\mathcal H}_{\mathrm{eff}}[\phi]}\equiv
{\int}{\mathcal{D}}\check{\psi}\,
{\mathrm e}^{-{\mathcal{H}}_2[\psi_{\mathrm{cl}}]-
\check{\mathcal{H}}[{\phi},\check{\psi}]-
{\mathcal{H}}'[{\phi},\check{\psi}]}\;.
\end{equation}
For ${\mathcal H}'=0$ we have
\begin{equation}\label{Hphi4eff}
{\mathcal H}_{\mathrm{eff}}[\phi]=
\frac{1}{2}\,\ln\frac{\check{b}_2}{2\pi}+
{\mathcal H}_1[\phi;\check{a}_2,a_4^{\mathrm{eff}}]
\end{equation}
where ${\mathcal H}_1[\phi;\check{a}_2,a_4^{\mathrm{eff}}]$ is the
$h=0$ variant of the Hamiltonian (\ref{H1}), with $a_2$ and $a_4$
replaced by expression (\ref{acheck}) for $\check{a}_2$ and
\begin{equation}
a_4^{\mathrm{eff}}
=a_4-\frac{d_{21}^2}{2\,\check{b}_2}\;,
\end{equation}
respectively. Thus, if ${\mathcal H}'$ can be set to zero, the critical behavior of the Hamiltonian (\ref{Hrewr}) reduces indeed to that of a $\phi^4$ model.
This result is, of course, not new: it has been used in the construction
and analyses of the
dynamic models C, D, and E for a long time (Cf.\ also Ref.\
 \cite{PPK73}).

Next, consider what happens when ${\mathcal H}'$ is included.
A straightforward calculation shows 
 that the cubic and quartic  terms of
${\mathcal H}'$ can be rewritten as
\begin{equation}\label{psi3rw}
{\int_\Omega}\,\check{\psi}^3=
{{\mathfrak{G}}_{\mathrm{tra}}[\Upsilon_3]}\,
\check{{\mathcal H}}[\phi,\check{\psi}]-
{\int_\Omega}{\left[
\frac{d_{21}^3}{8\,\check{b}_2^3}\,\phi^6
+{3\,d_{21}\,\delta(\boldsymbol{0})
\over 2\,\check{b}_2^2}\,\phi^2
\right]}
\end{equation}
and
\begin{eqnarray}\label{psi4rw}
{\int_\Omega}\,\check{\psi}^4&=&
{{\mathfrak{G}}_{\mathrm{tra}}[\Upsilon_4]}\,
\check{{\mathcal H}}[\phi,\check{\psi}]
+
{\int_\Omega}{\left[\frac{d_{21}^4}{16\,\check{b}_2^4}\,\phi ^8
\right.}\nonumber\\&&\mbox{}
+{3\,d_{21}^2\,{\delta(\boldsymbol{0})}\over
2\,\check{b}_2^3}\,\phi^4
+{\left.\frac{3\,{\left[\delta(\boldsymbol{0})\right]}^2}{\check{b}_2^2}
\right]}
\end{eqnarray}
with
\begin{equation}
\Upsilon_3={\check{\psi}^2\over \check{b}_2}-
{d_{21}\,\phi^2\over 2\,\check{b}_2^2}\,\check{\psi}
+{d_{21}^2\,\phi^4+
8\,\check{b}_2\,\delta(\boldsymbol{0})\over 4\,\check{b}_2^3}
\end{equation}
and
\begin{eqnarray}
\Upsilon_4&=&{\check{\psi}^3\over \check{b}_2}-
{d_{21}\phi^2\over 2\,\check{b}_2^2}\,\check{\psi}^2
+{d_{21}^2\phi^4+
12\,\check{b}_2\,\delta({\boldsymbol{0}})\over 4\,\check{b}_2^3}\,\check{\psi}
\nonumber\\&&\mbox{}
-{d_{21}\,\phi^2\over 8\, \check{b}_2^4}
  {\left[20\,\check{b}_2\,\delta(\boldsymbol{0}) + 
    d_{21}^2\,\phi^4\right]}\;.
\end{eqnarray}

The meaning of these results is the following.
To first order in the coupling constants $\check{b}_3$ and $b_4$,
the terms $\propto {\int_\Omega}\,\check{\psi}^3 $ and
$\propto{\int_\Omega}\,\check{\psi}^4$ can be transformed
away at the expense
of (i) additional irrelevant interactions ($\propto \phi^n$, with $n=6,8$), (ii)
contributions $\propto |\Omega|$, and (iii) a change of the coefficients of
the $\phi^2$ and $\phi^4$ terms.

The irrelevant interactions may be dropped. (iii) can be absorbed through
a redefinition of the temperature-like scaling field and the irrelevant
scaling field $\sim u-u^*$, which hence  become dependent on
$\check{b}_3$ and $b_4$ (as well as on
 $d_{21}$ and $\check{b}_2$). (ii) means contributions $\propto \check{b}_3$ and
$\propto b_4$ to the constant part of the Hamiltonian, which we write as
$\mu_0\,|\Omega|$, where $\mu_0$ is Wegner's \emph{ special} scaling field
(cf.\ Sec.\ III.G.2 of  Ref.\ \cite{Weg76}). Since $\check{b}_2$ depends via
$\psi_{\mathrm{cl}}$ on the nonordering field $g$, 
these contributions to $\mu_0$ 
have {\em $g$-dependent parts\/}.

We are now ready to understand the origin and nature of the discontinuity
eigenperturbation indicated in Fig.~\ref{fig:RGflpat}. Let
$\mu_0^{(\lambda)}(g)\equiv \mu_0[a_{2\lambda}(g),g]$ be the value of $\mu_0$
at a point $(a_{2\lambda}(g),h{=}0,g)$ on the $\lambda$-line%
\footnote{We suppress the dependence of $\mu_0$ and $a_{2\lambda}$ on the other
variables, $b_2$, $b_4$, etc. Instead of the variables $a_2$, \ldots
one could, of course, also use $\check{a}_2$, \ldots.
Yet we prefer  to express $\mu_0$ in terms
of the former.}%
, and
$\mu_0^{({\mathrm{e}})}\equiv\mu_0^{(\lambda)}(g_{\mathrm{e}})$,
 the corresponding critical-end-point value.
Consider a variation $g_{\mathrm{e}}\to g_{\mathrm{e}}+\delta g$,
$a_{2{\mathrm{e}}}\to
% a_{2\lambda}(g_{\mathrm{e}}+\delta g)=
a_{2{\mathrm{e}}}+\delta a_2$
along the $\lambda$-line. This induces a change
\begin{equation}\label{dep}
\delta\mu_0^{(\lambda)}=\mu_0^{(\lambda)}(g)-\mu_0^{({\mathrm{e}})}
\end{equation}
of the special scaling field $\mu_0$, with 
$\delta\mu_0^{(\lambda)}\sim
T_{\mathrm{c}}(g)-T_{\mathrm{e}}$ for small $\delta g$.
The shift of the Hamiltonian
$\delta\mu_0^{(\lambda)}\,|\Omega|$ is an eigenperturbation
with eigenexponent $y=d$. Obviously this is the one we were looking for
and whose eigendirection is shown in
Fig.~\ref{fig:RGflpat}.

We still have to consider the $(\nabla\check{\psi})^2$ term of ${\mathcal{H}}'$.
Power counting tells us that it is irrelevant.
One also verifies that it can be rewritten as
\begin{eqnarray}\label{gradpsi2rw}
{\int_\Omega}\,(\nabla\check{\psi})^2&=&
{{\mathfrak{G}}_{\mathrm{tra}}[-\check{b}_2^{-1}\triangle\check{\psi}]}\,
\check{{\mathcal H}}[\phi,\check{\psi}]\nonumber\\&&\mbox{}
+\check{b}_2^{-1}
{\int_\Omega}{\left[\frac{d_{21}}{2}\,\phi^2\triangle\check{\psi}
+\int_{\boldsymbol{q}}q^2\right]}.\qquad
\end{eqnarray}
The last term in the second line contributes again to the constant part $\mu_0\,|\Omega|$ of the Hamiltonian. The first one, proportional to
$\int_{\Omega}\phi^2\,\triangle\psi$,
is irrelevant according to power counting, which conforms with
our expectation that the interaction (\ref{gradpsi2rw}) is irrelevant
(apart from contributions to the constant part of the Hamiltonian)
and can be dropped.
Finally, power counting also indicates that
the terms involving $e_{21}$ and $f_{21}$ are irrelevant, so that they
may be omitted as well.

Having identified the sought discontinuity eigenperturbation (\ref{dep})
and knowing that it does not contribute
to the critical singularities on the $\lambda$-line or at the end point,
we can now focus directly on these, dropping the
$\check{\psi}^3$ and $\check{\psi}^4$ interactions as well as the
$(\nabla\check{\psi})^2$ term (\ref{gradpsi2rw}).
From Eqs.~(\ref{Hphieffdef}) and (\ref{Hphi4eff}) we know already that
the resulting truncated Hamiltonian $\check{\mathcal{H}}$ reduces to an effective $\phi^4$
Hamiltonian upon integrating out $\check{\psi}$.
Moreover, the Hamiltonian $\check{\mathcal{H}}$ and its renormalization
are well known from studies of the stochastic dynamic models C, D, and E
(see, e.g., Refs.~\cite{BdD75}, \cite{ES80},  or Sec.\ 35.4 of \cite{ZJ96}).
Adding a source term $-{\int_\Omega}  \check{I}\,\check{\psi}$ to $\check{\mathcal{H}}$,
one can compute the functional integral
${\int}{\mathcal{D}}\check{\psi}\;
{\exp}{\left\{-\check{\mathcal{H}}[{\phi},\check{\psi}]
+{\int_\Omega} \check{I}\check{\psi}\right\}}$ 
to relate the cumulants
$\langle \prod_{j=1}^N\phi({\boldsymbol{x}}_j)\,\prod_{i=1}^M\psi
({\boldsymbol{X}}_i)
\rangle^{\mathrm{cum}}_{\check{\mathcal{H}}}$ pertaining to
$\check{\mathcal{H}}$ to cumulants
of a standard $\phi^4$ theory described by the Hamiltonian (\ref{Hphi4eff}).
One has
\begin{equation}\label{psiphi2}
\langle\check{\psi}({\boldsymbol{X}})
\rangle_{\check{\mathcal{H}}}=-
{d_{21}\over 2\check{b}_2}\,\langle\phi^2({\boldsymbol{X}})
\rangle_{{\mathcal{H}}_{\mathrm{eff}}}
\end{equation}
and
\begin{equation}
\langle\check{\psi}({\boldsymbol{X}}_1)\,\check{\psi}({\boldsymbol{X}}_2)
\rangle^{\mathrm{cum}}_{\check{\mathcal{H}}}
={\delta({\boldsymbol{x}}_{12})\over \check{b}_2}+
{d_{21}^2\over  4\,\check{b}_2^2}\,\langle\phi^2({\boldsymbol{X}}_1)\,
\phi^2({\boldsymbol{X}}_2)
\rangle^{\mathrm{cum}}_{{\mathcal{H}}_{\mathrm{eff}}}\;.
\end{equation}
All other correlation functions  of the fields
$\check{\psi}$ and $\phi$
are identical to their analogs of the effective $\phi^4$ theory
one obtains through the replacement
\begin{equation}
\check{\psi}({\boldsymbol{X}})\to 
-{d_{21}\over 2\check{b}_2}\,\phi^2({\boldsymbol{X}})\;.
\end{equation}
Thus the critical behavior of  the above cumulants
$\langle \prod_{j=1}^N\phi({\boldsymbol{x}}_j)\,\prod_{i=1}^M\psi
({\boldsymbol{X}}_i)
\rangle^{\mathrm{cum}}_{\check{\mathcal{H}}}$
 may be inferred directly from
the known one of the corresponding cumulants of the $\phi^4$ theory.
Alternatively, one
could derive RG equations for the former, using the
reparametrizations $a_4^{\mathrm{eff}}=\kappa^\epsilon\,Z_u(u)\,u$,
$\phi=Z^{1/2}_\phi(u)\,\phi_{\mathrm{ren}}$,
$\check{\psi}=\check{b}_2^{-1/2}\,
Z^{1/2}_\psi(u,\gamma)\,\check{\psi}_{\mathrm{ren}}$,
$a_{2}^{\mathrm{eff}}=\kappa^2\,Z_{\tau}(u)\,\tau+a^{\mathrm{eff}}_{2,c}$,
$\check{a}_2=\kappa^2\,Z_{\psi}(u,\gamma)\,Z_\tau(u)\,\check{\tau}+\check{a}_{2,c}$,
and $d_{21}=\check{b}_2^{1/2}\,\kappa^{\epsilon/2}\,
Z^{1/2}_\psi(u,\gamma)\,
Z_{\tau}(u)\,\gamma$.
Here $Z_u$, $Z_\phi$, and $Z_\tau$ are the
renormalization factors introduced in Eqs.~(\ref{DefZfactors})--(\ref{hren}),
while $Z_\psi$ is of the form $Z_\psi(u,\gamma)=1/[1-\gamma^2\,f(u)]$,
where $f(u)$ is known to be
given by the Laurent part of the additive counterterm  $\propto I^2$
in Eq.~(\ref{genfctG1}) \cite{BdD75,ES80}.

From Eq.~(\ref{genfctG1}) we see that the leading singularity of 
$\langle\check{\psi}\rangle$ (and hence of $\langle\psi\rangle$)
on a path with $\tau\sim T-T_c(g)$ agrees with that of the energy
density $\langle\phi^2\rangle$. Since on the liquid branch of the
liquid-gas coexistence boundary the scaling field
$\tau$ varies asymptotically
linearly in $t\sim T{-}T_{\mathrm{e}}$ as the critical end point is approached,
the leading singularity of the density $\varrho_{\mathrm{tot}}=\langle\psi\rangle$
is indeed of the form $V^0_\pm\,\left|{t}\right|^{1-\alpha}$, where
$V^0_+/V^0_-$ is the usual universal specific-heat amplitude ratio.

\subsubsection{Nonsymmetric case}{\label{sec:CoupThnsym}}

Considering points $(a_2,h,g)$
on the $\beta\gamma$ (disordered) side
of the coexistence surface ${\boldsymbol{\rho}}$,
we expand about the classical solutions
$\phi_{\beta\gamma}^{\mathrm{cl}}$ and
$\psi_{\beta\gamma}^{\mathrm{cl}}$.
Since we ultimately wish to approach
the  $\lambda$-line or critical end point,
we may restrict ourselves to
points in their immediate vicinity.
Introducing the shifted densities $\check{\psi}$ and $\check{\psi}$ via
\begin{equation}
\phi=\phi_{\mathrm{cl}}+\check{\phi}\;,\quad\psi=\psi_{\mathrm{cl}}+\check{\psi}\;,
\end{equation}
we express the Hamiltonian defined through Eqs.~(\ref{ham})
and (\ref{H1})--(\ref{H12})
in terms of these.
The resulting quadratic (`Gaussian') part can be written as
\begin{eqnarray}
{\mathcal{H}}_{\mathrm{G}}&=&
\frac{1}{2}{\int_\Omega}
\left(\check{\phi}\,,\;\check{\psi}\right)
\left[ 
\boldsymbol{\mathcal{V}}^{(2)}_{\mathrm{cl}}
-\boldsymbol{\eta}_{\mathrm{cl}}\,\triangle
\right]
\left(\begin{array}{c}
\check{\phi}\\
\check{\psi}\end{array}\right)
\end{eqnarray}
with
\begin{eqnarray}
\boldsymbol{\eta}_{\mathrm{cl}}\equiv
\left(\begin{array}{c@{\;\;}c}
A+f_{21}\,\psi_{\mathrm{cl}}&e_{11}+e_{21}\,\phi_{\mathrm{cl}}\\[0.1em]
e_{11}+e_{21}\,\phi_{\mathrm{cl}}&B
\end{array}\right)
\end{eqnarray}
and $\boldsymbol{\mathcal{V}}^{(2)}_{\mathrm{cl}}
\equiv\boldsymbol{\mathcal{V}}^{(2)}(\phi_{\mathrm{cl}},\psi_{\mathrm{cl}})$,
where  $\boldsymbol{\mathcal{V}}^{(2)}$  is the matrix introduced in Eq.~(\ref{V2}).

The matrix $\boldsymbol{\eta}_{\mathrm{cl}}$ must be positive definite. [Otherwise,
the classical homogeneous states $(\phi_{\mathrm{cl}},\psi_{\mathrm{cl}})$
are not  guaranteed to be  stable.] It plays the role of a metric.
Hence we can diagonalize
${\mathcal{H}}_{\mathrm{G}}$ by a similarity transformation
\begin{eqnarray}
\left(\begin{array}{c}
\check{\phi}\\
\check{\psi}\end{array}\right)
&=&{\boldsymbol{U}}\cdot
\left(\begin{array}{c}
\varphi_1\\
\varphi_2\end{array}\right)\;,
\end{eqnarray}
satisfying the diagonalization condition (\ref{diag}) together with the orthonormality
relation
\begin{equation}\label{ONrel}
{\boldsymbol{U}}^{\mathrm{T}}\cdot\boldsymbol{\eta}_{\mathrm{cl}}
\cdot{\boldsymbol{U}}={\boldsymbol{1}}\;,
\end{equation}
to obtain
\begin{equation}\label{HGdiag}
{\mathcal{H}}_{\mathrm{G}}=\frac{1}{2}{\int_\Omega} \!
{\left[(\nabla\varphi_1)^2+\lambda_1\,\varphi_1^2+
(\nabla\varphi_2)^2+\lambda_2\,\varphi_2^2
\right]}\;.
\end{equation}
By analogy with Eqs.~(\ref{Hrewr})--(\ref{Hprime}),
 we split the resulting total Hamiltonian (\ref{ham}) as
\begin{equation}
{\mathcal{H}}[\phi,\psi]={\mathcal{H}}[\phi_{\mathrm{cl}},\psi_{\mathrm{cl}}]+
\check{\mathcal{H}}[\varphi_1,\varphi_2]+{\mathcal{H}}'[\varphi_1,\varphi_2]
%+{\mathcal{H}}''[\varphi_1,\varphi_2]
\end{equation}
with
\begin{eqnarray}
\check{\mathcal{H}}[\varphi_1,\varphi_2]&=&
{\int_\Omega}{\left[
\frac{1}{2}\left(\nabla\varphi_1\right)^2
+{\lambda_1\over 2}\,\varphi_1^2
+{v_{4,0}\over 4}\,\varphi_1^4
\right.}\nonumber\\&&{\left.\mbox{}\qquad
+{\lambda_2\over 2}\,\varphi_2^2
+\frac{v_{2,1}}{2}\,\varphi_1^2\,\varphi_2
\right]}
\end{eqnarray}
and
\begin{eqnarray}\label{Hprimenew}
{\mathcal{H}}'[\varphi_1,\varphi_2]&=&
{\int_\Omega}\bigg[\frac{1}{2}\left(\nabla\varphi_2\right)^2
+\sum_{j=3}^4\frac{v_{0,j}}{j}\,\varphi_2^j
\nonumber\\&&\mbox{}
+v_{1,2}\,\varphi_1\,\varphi_2^2
+v_{3,1}\,\varphi_1^3\,\varphi_2
+v_{1,3}\,\varphi_1\,\varphi_2^3
\nonumber\\&&\mbox{}
+v_{2,2}\,\varphi_1^2\,\varphi_2^2+
\frac{e_{21}}{2}\,\check{\phi}^2\triangle\check{\psi}
+\frac{f_{21}}{2}\,{\left(\nabla\check{\phi}\right)}^2\psi
\bigg].\nonumber\\
\end{eqnarray}

We have dropped a contribution $v_{3,0}\,\varphi^3$ to the integrand
of ${\mathcal{H}}'$; its coupling constant $v_{3,0}$ vanishes
at the Landau-theory location of the $\lambda$-line by
condition (\ref{v30}), and we know already that it can be transformed away
by means of a shift $\varphi_1\to\varphi_1+\Phi_1$; i.e.,
aside from a change of the interaction constants
of the Hamiltonian $\check{\mathcal{H}}$ and its constant part,
it corresponds to a sum of redundant and irrelevant operators.

The Hamiltonian $\check{\mathcal{H}}$ has the familiar model-C form
(\ref{Hcheck}). The contributions in the first line of Eq.~(\ref{Hprimenew})
are the analogs of the Hamiltonian (\ref{Hprime}); from our considerations in
Sec.~\ref{sec:CoupThsym} we know their effect: Apart from contributing to the
special scaling field (and shifting the $\lambda$-line and critical end point), they
involve redundant and irrelevant operators, and hence may be dropped when
analyzing the asymptotic critical behavior. The same is true of the
remaining terms of (\ref{Hprime}). The easiest way
to arrive at this conclusion is
via  power counting, utilizing the canonical scale dependences
 $\varphi_1\sim \kappa^{1-\epsilon/2}$ and $\varphi_2\sim
\kappa^{2-\epsilon/2}$. Moreover, the contributions involving the interaction
constants $v_{1,2}\,,\ldots,v_{2,2}$ could be analyzed along the same lines as
the $\check{\psi}^3$ and $\check{\psi}^4$ nonlinearities in the previous subsection
by  choosing appropriate functionals
$\Upsilon[\varphi_2]$ to eliminate them via
a change of variable $\varphi_2\to\varphi_2+\Upsilon[\varphi_2]$.
The contributions  proportional to $e_{21}$ and $f_{21}$ correspond to a variety of
derivative terms $\varphi_1^2\triangle\varphi_1$, $\varphi_1^2\triangle\varphi_2$,
\ldots, $(\nabla\varphi_2)^2\varphi_2$, all of which are irrelevant
according to power counting.

Upon ignoring ${\mathcal{H}}'$ altogether, we get back to the Hamiltonian $\check{\mathcal{H}}[\varphi_1,\varphi_2]$, which is equivalent
to the one in Eq.~(\ref{Hcheck}). The implications
for the asymptotic critical behavior that occurs
when the $\lambda$-line or the critical end point
are approached is obvious: $\varphi_1$, the order parameter,
and the  secondary density $\varphi_2$ behave asymptotically as
their respective analogs $\phi$ and $\psi$ in the symmetric case.

The density $\psi$ has components along both $\varphi_1$ and
$\varphi_2$. Consequently its average
$\varrho_{\mathrm{tot}}=\langle\psi\rangle$
has singularities of the form
$U^0_\pm\,\left|{t}\right|^{\beta}$ (with $U^0_+\equiv 0$) and $V^0_\pm\,\left|{t}\right|^{1-\alpha}$ in the limit $t\equiv(T-T_{\mathrm{e}})/T_{\mathrm{e}}\to \pm 0$.
This holds, in particular,
for approaches along the coexistence boundary $g_\sigma(t)$.
Thus the leading singularity of
the liquid density $\varrho_{\mathrm{tot}}$ is $\sim |t|^\beta$
if the critical end point is approached along the triple line,
and $\sim |t|^{1-\alpha}$ if it is approached along
the $\beta\gamma$-$\alpha$ section of the liquid-gas coexistence line.
These findings are in conformity with Eq.~(\ref{singdens}).

\subsubsection{The singularity of the first-order line $g_\sigma(t)$}
{\label{sec:Sing1ol}}

A crucial assumption on which the phenomenological derivation
\cite{Fis89}
of the $|t|^{2-\alpha}$ singularity of
the first-order line $g_\sigma(t)$ is based is the asymptotic behavior of
the grand potentials ${\mathcal{A}}_\wp$ of the phases $\wp=\beta$,
$\gamma$, and $\beta\gamma$ at the critical end point.
This is hypothesized to be of the same form as at a conventional critical point; i.e.,
it can be decomposed by analogy with Eq.~(\ref{fregsing})
into a regular background contribution
and a nonanalytic part, ${\mathcal{A}}_\wp^{\mathrm{sing}}$,
which has the asymptotic scaling form (\ref{fsing}).

Keeping in mind our identification (\ref{dep}) of the discontinuity
eigenperturbation, one realizes that our
above field-theoretic considerations are in full accordance with
these assumptions, corroborating them.
Utilizing them one can proceed \cite{Dr-MS99} just as in the
derivation \cite{Fis89}
based on the phenomenological scaling theory
to determine the coexistence singularity
of the first-order boundary $g_\sigma(t)$ from the
coexistence conditions (\ref{coexcond}). Since $\hat{\tau}\sim t$
for small deviations from the critical end point directed along
$g_\sigma(t)$, this line inherits the $|t|^{2-\alpha}$ singularity
of the grand potentials ${\mathcal{A}}_{\beta\gamma}$ and ${\mathcal{A}}_{\beta,\gamma}$. Thus Eq.~(\ref{singpb}) follows.

\section{Summary and conclusion}\label{sec:concl}

To put things in perspective, it will be helpful to recapitulate
the main steps of our analysis and summarize its principal results.

(i) In order to study the critical behavior
at symmetric and nonsymmetric critical end points
as well as the associated coexistence singularities,
phenomenological continuum models with a Hamiltonian
depending on two fluctuating densities $\phi$ and $\psi$
were introduced. These models may be viewed as continuum
variants of the BEG model \cite{BEG71}. They generalize
similar ones previously considered in the literature
(cf.\ Ref.~\cite{Wil97a}), to which they reduce when
some interaction constants vanish. We justified them
via phenomenological arguments, but were also able to derive
them directly from the BEG model, utilizing a generalized
Kac-Hubbard-Stratonovich transformation to map the latter
exactly on a lattice field theory and a subsequent continuum
approximation.

(ii) We then employed Landau theory to analyze the models for
the cases in which the Hamiltonian obeys or breaks the symmetry
$\phi\to-\phi$. Provided the interactions constants were chosen
in the appropriate ranges, phase diagrams with either a symmetric
or nonsymmetric critical end point could be obtained
(cf.\ Figs.~\ref{fig:pdsmixed}--\ref{fig:pdns}).
Furthermore, we corroborated that the $|t|^{2-\alpha}$ singularity (\ref{singpb})
of the first-order line $g_\sigma(t)$ and the
$|t|^{1-\alpha}$ and $|t|^\beta$ coexistence singularities (\ref{singdens})
of $\langle\psi\rangle$ take on the expected Landau-theory forms, namely,
discontinuities in the second derivative of  $g_\sigma(t)$ and
in the first derivative of  $\langle\psi\rangle$, or
a singularity $\sim |t|^{1/2}$, respectively.

(iii) Building on these results, we then studied the critical behavior on the
$\lambda$-line and at the critical end point.
One important observation was the following. In order to obtain phase
diagrams with a critical end point (in Landau theory and beyond),
it was \emph{essential to include nonlinearities}
such as a $\psi^4$ term
\emph{that are irrelevant according to power counting}.
The presence of  this term
guarantees that a stable $\alpha$ (gas) phase
exists, even in the absence of coupling to the (primary) order-parameter density
$\phi$. However, once it has been verified that
the phase diagram has the correct topology (with a critical line $\lambda$,
critical end point, and first-order boundary $g_\sigma(t)$),
one can focus on the more specific problem of the asymptotic critical behavior
that occurs when either a point on the $\lambda$-line or else the critical end point itself is approached along a given thermodynamic path. Then approximations that are
tailored for this particular purpose become viable.

Upon expanding about the classical equilibrium
values of the densities $\phi$ and $\psi$, and expressing
the Hamiltonian in terms of the deviations
$\check{\phi}$ and $\check{\psi}$ from these,
the Hamiltonian was found to involve also nonlinearities
 that are \emph{irrelevant} according
to power counting,
such as the $\check{\psi}^3$ and $\check{\psi}^4$
terms in the symmetric case.
We were able to show, within the context of perturbation theory,
what their effects are: Aside from (a) inducing
 \emph{a change of the usual  two scaling fields}
(the temperature-like scaling field $\tau$ and
the magnetic-field like scaling field),
they (b) correspond to \emph{irrelevant  operators}
(giving rise to corrections to scaling)
and (c)  contribute to the \emph{constant part of
the fixed-point Hamiltonian}, i.e.,
to Wegner's \cite{Weg76} \emph{special} scaling field $\mu_0$.

(iv) These findings enabled us to verify the existence of
a discontinuity eigenexponent $d$ at the
fixed point representing the critical end point, and to understand
its origin: The eigenperturbation with which it is associated corresponds
to the change $\delta\mu_0$ of the special scaling field $\mu_0$ that occurs
when one moves away from the critical end point such that the theory
remains critical.

(v) In conjunction with the findings mentioned at the end of (iii), (iv)
shows that the critical behavior on the $\lambda$-line is
the same as at the critical end point, inasmuch as
the values of the critical exponents and other universal quantities
are concerned.

We emphasize that our conclusions mentioned in (iv) and (v)
are in conformity with the
position-space RG results of Ref.~\cite{BW76,KGYF81} and
the RG picture that has emerged from them (cf.\ Fig.\ \ref{fig:RGflpat}).
Our results show how this RG picture translates into field theory;
they verify the existence of the discontinuity eigenexponent, clarify its
origin, and reveal how its presence can be reconciled with
 the anticipated equality of the critical exponents
associated with the critical line and the critical end point, respectively.

(vi) Once we had established the results (iv) and (v),
we could derive the coexistence singularities (\ref{singpb}) and
(\ref{singdens}) in a fairly straightforward fashion.
The $|t|^{1-\alpha}$ singularities in Eq.~(\ref{singpb})
are due to the coupling of the secondary density, $\psi$, to
the energy density. In the case of a \emph{symmetric} critical end point,
this is the leading temperature singularity of $\langle\psi\rangle$
on the liquid branch of the liquid-gas coexistence curve.
In the case of a \emph{nonsymmetric} critical end point,
the densities $\phi$ and $\psi$ mix. Thus the order parameter
becomes a linear combination, $\varphi_1$, of both, so that $\psi$
couples directly to $\varphi_1$. This implies that the leading
coexistence singularity of $\langle\psi\rangle$ on the
three-phase-coexistence side of the transition becomes $\sim|t|^\beta$. 
The $|\tau|^{2-\alpha}$ of the first-order boundary $g_\sigma(t)$
follows in a way known from its original phenomenological derivation \cite{Fis89}
from the usual free-energy singularity of the $\beta\gamma$, $\beta$, and $\gamma$
phases at the critical end point.

(vii) A challenging issue raised by
Fisher and Barbosa  in their critical assessment \cite{FB91}
of the theory of critical end points, is the potential occurrence of essential
singularities at  first-order boundaries like $\boldsymbol{\sigma}$.
Since our approach relied on perturbative RG arguments,
essential singularities are beyond its scope.

The following should also be clear: We  cannot, of course, rule out that
for special models the critical behavior at the critical end point might
differ from that on the critical line. However, this should not happen in the
generic cases we were concerned with here.

(viii) The present work, which was focused exclusively
on \emph{bulk} properties, provides a basis for investigating
the problem of critical adsorption at  a noncritical $\alpha$-$\beta\gamma$ interface.
This issue will be taken up in a separate publication.

(ix) We close with a comment on possible implications of the
recent work by Fisher \emph{et al} \cite{FO00,OFU00} that appeared
after completion of the present investigation. Upon reanalyzing
two-phase heat-capacity data of the one-component fluid
propane (C$_3$H$_8$), these authors
concluded that both contributions to the specific heat
\begin{equation}
C_V^{\mathrm{tot}}=VT(\partial^2p/\partial T^2)_V-
NT(\partial^2\mu/\partial T^2)_V\;,
\end{equation}
the pressure derivative $d^2p/dT^2\equiv p''_{\mathrm{vp}}$ \emph{and}
the chemical potential derivative $d^2\mu/dT^2\equiv \mu''_{\mathrm{vp}}$
diverge as $|t|^{-\alpha}$ when $t\equiv (T-T_c)/T_c\to -0$ on the
vapor-pressure curve. This behavior, which Yang and
Yang \cite{YY64} originally suggested to be the most likely one
for real gases, differs from that of simple
lattice gas models, for which $\mu_{\mathrm{vp}}$ is found to be analytic
through $T_c$ so that $\mu''_{\mathrm{vp}}(T_c-)$ must remain finite.
Building  on Rehr and Mermin's earlier work \cite{RM73}, Fisher and
Orkoulas \cite{FO00} concluded that the standard scaling theory
for fluid criticality must be revised inasmuch as the pressure difference
$p-p_c$ should, in general, also mix into the scaling fields, in addition
to the chemical potential and temperature differences $\mu-\mu_c$ and $t$,
respectively. As a remarkable consequence of this `pressure mixing'  they
found that the arithmetic mean $(\rho_{\mathrm{l}}+\rho_{\mathrm{g}})/2$ of the
gas and liquid densities $\rho_{\mathrm{l,g}}$ at coexistence should have
a temperature singularity  $\sim |t|^{2\beta}$ as $t\to -0$.
Such a singularity would dominate over the usual
 energy-density singularity $\sim |t|^{1-\alpha}$ that is usually given in textbooks
\cite{CL95} as the leading one of this quantity.

If this pressure mixing  must indeed be taken into account in the analysis of critical
behavior of simple fluids, one anticipates it to play a role also in the case of
critical behavior at nonsymmetric critical end points. Clearly, the
leading singularity $\sim |t|^\beta$ of the coexistence density (\ref{singdens})
must prevail, but one expects pressure mixing to produce
a subleading one of the form $|t|^{2\beta}$, which would be stronger
than the subleading singularity $\sim |t|^{1-\alpha}$  
included in Eq.~(\ref{singdens}).

\begin{acknowledgement}
One of us (H.W.D.) is indebted to Royce K.\ P.\ Zia
for helpful correspondence, discussions, and a critically reading of the
manuscript. The other author (M.S.) would like
to thank P.\ Upton and J.\ Eggers for helpful discussions.We also thank Boris N.\ Shalaev for
enjoyable discussions and Ralph Blossey for taking the time to carefully
read the manuscript.
This work has been supported through the Deutsche
Forschungsgemeinschaft (DFG)  via Sonderforschungsbereich 237
and the Leibniz program (Di387/2-1).
\end{acknowledgement}

\appendix
\section{Mapping of the BEG model onto a field theory}\label{sec:app1}

We start from the Hamiltonian (\ref{HBEG}) of the BEG model.
To set up the Hubbard-Stratonovich transformation, we introduce the row vectors
\begin{eqnarray}
\boldsymbol{\sigma}=\left(\ldots\,,\, S_{\boldsymbol{i}}\,,\,\ldots
,\,S_{\boldsymbol{j}}^2\,,\ldots
\right),
\end{eqnarray}
\begin{equation}
\boldsymbol{F}=\left(\ldots,\, H_{\boldsymbol{i}}=H\,,\,\ldots,\,
D_{\boldsymbol{j}}=D\,,\,\ldots
\right),
\end{equation}
the transposed column vector $\boldsymbol{\sigma}^\mathrm{T}$,
and the matrix
\begin{equation}
\boldsymbol{M}=\left(\begin{array}{cc}
\boldsymbol{J}+M_0\,\boldsymbol{1}&\boldsymbol{L}\\\boldsymbol{L}&
\boldsymbol{K}+M_0\,\boldsymbol{1}\
\end{array}\right)\,.
\end{equation}
Here $\boldsymbol{J}=(J_{\boldsymbol{i}\boldsymbol{j}})$, $\boldsymbol{K}=(K_{\boldsymbol{i}\boldsymbol{j}})$,
and $\boldsymbol{L}=(L_{\boldsymbol{i}\boldsymbol{j}})$ are the matrices of interaction constants,
i.e., $J_{\boldsymbol{i}\boldsymbol{j}}$ takes the value $J$ or vanishes, depending on whether
$\boldsymbol{i}$
and $\boldsymbol{j}$ are nearest neighbors or not, and likewise for $\boldsymbol{L}$ and $\boldsymbol{K}$. The parameter $M_0$ has been chosen such that
the matrix $\boldsymbol{M}$ is positive definite (cf.\ \cite{Fis83a} and \cite{DC91}),
so that the inverse of $\boldsymbol{M}$ exists.

With these definitions the Hamiltonian (\ref{HBEG}) can we written
as 
\begin{equation}
{\mathcal{H}}_\mathrm{BEG}=-\frac{1}{2}\,\boldsymbol{\sigma}
\cdot\left(\boldsymbol{M}-M_0\right)\cdot
\boldsymbol{\sigma}^\mathrm{T}-\boldsymbol{F}\cdot \boldsymbol{\sigma}^\mathrm{T}\;.
\end{equation}
Proceeding in a standard fashion by making a Gaussian transformation,
we can perform the trace over the spin variables in the partition function
(\ref{ZBEG}). This gives
\begin{equation}
{\mathcal{Z}}_\mathrm{BEG}=\mathrm{e}^{-f_0}\left(\prod_{\boldsymbol{i}}
{\int_{-\infty}^\infty}\!d\phi_{\boldsymbol{i}}
{\int_{-\infty}^\infty}\!d\psi_{\boldsymbol{i}}\right)
\mathrm{e}^{-{\mathcal{H}}_\mathrm{lft}[\phi,\psi]}
\end{equation}
with
\begin{equation}
f_0= \frac{1}{2}\,\mathrm{Tr}\ln\left[2\pi\,(\boldsymbol{M}-M_0)\right]
\end{equation}
and
\begin{eqnarray}\label{Hlft}
{\mathcal{H}}_\mathrm{lft}
%[\boldsymbol{\phi},\boldsymbol{\psi}]
&=&
\frac{1}{2}\sum_{\boldsymbol{i}\ne\boldsymbol{j}}\left[\phi_{\boldsymbol{i}}\,
{\mathcal{P}}_{\boldsymbol{i}\boldsymbol{j}}\,\phi_{\boldsymbol{j}}+
\psi_{\boldsymbol{i}}\,
{\mathcal{Q}}_{\boldsymbol{i}\boldsymbol{j}}\,\psi_{\boldsymbol{j}}
%\right.\nonumber\\&&\mbox{}\left.
+{\mathcal{R}}_{\boldsymbol{i}\boldsymbol{j}}\left(\phi_{\boldsymbol{i}}\,\psi_{\boldsymbol{j}}+
\psi_{\boldsymbol{i}}\,\phi_{\boldsymbol{j}}\right)\right]
\nonumber\\&&\mbox{}
+\sum_{\boldsymbol{i}}w(\phi_{\boldsymbol{i}},\psi_{\boldsymbol{i}})\,,
\end{eqnarray}
where
\begin{eqnarray}
w(\phi_{\boldsymbol{i}},\psi_{\boldsymbol{i}})&=&
\frac{1}{2}\left(
{\mathcal{P}}_{\boldsymbol{i}\boldsymbol{i}}\,\phi_{\boldsymbol{i}}^2
+2\,{\mathcal{R}}_{\boldsymbol{i}\boldsymbol{i}}\,\phi_{\boldsymbol{i}}\,\psi_{\boldsymbol{i}}
+{\mathcal{Q}}_{\boldsymbol{i}\boldsymbol{i}}\,\psi_{\boldsymbol{i}}^2
\right)
\nonumber\\&&\mbox{}
-\ln\!\left[1+2\,\mathrm{e}^{\psi_{\boldsymbol{i}}+D-M_0}
\cosh\!\left(\phi_{\boldsymbol{i}}+H\right)\right],
\end{eqnarray}
while
${\mathcal{P}}_{\boldsymbol{i}\boldsymbol{j}}$,
${\mathcal{Q}}_{\boldsymbol{i}\boldsymbol{j}}$, and 
${\mathcal{R}}_{\boldsymbol{i}\boldsymbol{j}}$ are the  elements of
the matrices
\begin{eqnarray}\label{Qmatrix}
\boldsymbol{\mathcal{P}}&=&\left[\boldsymbol{J}+
M_0-\boldsymbol{L}\left(\boldsymbol{K}+M_0\right)^{-1}\boldsymbol{L}\right]^{-1}
\nonumber\\&=&{1\over M_0}\left[\mathbf{1}-\frac{\boldsymbol{J}}{M_0}
+\frac{\boldsymbol{J}^2+\boldsymbol{L}^2}{M_0^2}+\ldots\right]\,,
\end{eqnarray}
\begin{eqnarray}\label{Rmatrix}
\boldsymbol{\mathcal{Q}}&=&\left[\boldsymbol{K}+
M_0-\boldsymbol{L}\left(\boldsymbol{J}+M_0\right)^{-1}\boldsymbol{L}
\right]^{-1}
\nonumber\\&=&{1\over M_0}\left[\mathbf{1}-\frac{\boldsymbol{K}}{M_0}
+\frac{\boldsymbol{K}^2+\boldsymbol{L}^2}{M_0^2}+\ldots\right]\,,
\end{eqnarray}
and
\begin{eqnarray}\label{Smatrix}
\boldsymbol{\mathcal{R}}&=&
-\left[\boldsymbol{J}+M_0\right]^{-1}\,\boldsymbol{L}
\left[\boldsymbol{K}+
M_0-\boldsymbol{L}\left(\boldsymbol{J}+M_0\right)^{-1}\boldsymbol{L}\right]^{-1}
\nonumber\\&=& \frac{1}{M_0}
\left[-\frac{\boldsymbol{L}}{M_0}
+\frac{\boldsymbol{LJ}+\boldsymbol{KL}}{M_0^2}+\ldots\right]\,,
\end{eqnarray}
respectively. Thus the BEG model (\ref{HBEG}) is exactly equivalent
to a lattice-field theory with Hamiltonian (\ref{Hlft}).

Just as in the simpler Ising case \cite{Fis83a,DC91}, the
bonds of this Hamiltonian (i.e., the off-diagonal
elements  $-{\mathcal{P}}_{\boldsymbol{i}\boldsymbol{j}}$,
$-{\mathcal{Q}}_{\boldsymbol{i}\boldsymbol{j}}$, and 
$-{\mathcal{R}}_{\boldsymbol{i}\boldsymbol{j}}$)
extend beyond nearest neighbors (NN).
From the expansions given in the second line of (\ref{Qmatrix})--(\ref{Smatrix})
one sees that the NN couplings are given by $\boldsymbol{J}/M_0^2$,
$\boldsymbol{K}/M_0^2$, and $\boldsymbol{L}/M_0^2$, respectively, up to
corrections that are smaller by a factor $M_0^{-1}$. Likewise,
next-nearest-neighbor couplings
are down by this factor, compared to the NN bonds.

In our interpretation of the BEG model as a model for binary fluids,
interchanging A and B particles corresponds to the transformation
$S_{\boldsymbol{i}}\to -S_{\boldsymbol{i}}$ for all sites $\boldsymbol{i}$.
Consequently, ${\mathcal{H}}_\mathrm{BEG}$ is AB-symmetric
if $H=L=0$. In this case, ${\mathcal{H}}_\mathrm{lft}$ evidently is \emph{even} in $\phi$,
as it must.
  
So far our reformulation of the BEG model has been exact. We now make
a continuum approximation. Introducing
smoothly interpolating fields $(\phi_\upsilon(\boldsymbol{x}))
\equiv(\phi(\boldsymbol{x}), \psi(\boldsymbol{x}))$,
we write $\phi^\upsilon_{\boldsymbol{i}}=\phi_\upsilon(a\boldsymbol{i})$,
where $a$ is the lattice constant, and use the expansion
\begin{equation}
\phi^\upsilon_{\boldsymbol{i}}-\phi^\upsilon_{\boldsymbol{j}}=
\boldsymbol{r}_{\boldsymbol{i}\boldsymbol{j}}{\cdot}\nabla
\phi_\upsilon\left[a\,(\boldsymbol{i}+\boldsymbol{j})/ 2\right]+\ldots
\end{equation}
for small $\boldsymbol{r}_{\boldsymbol{i}\boldsymbol{j}}=a\,(\boldsymbol{i}-\boldsymbol{j})$.
(To simplify our analysis, we take the sites $\boldsymbol{i}$
to be those of a simple $d$ dimensional cubic lattice with periodic
boundary conditions.) 

In this manner we obtain from ${\mathcal{H}}_\mathrm{lft}$ the Hamiltonian
\begin{eqnarray}\label{cft1}
{\mathcal{H}}[\phi,\psi]&=&{\int_\Omega}\!d^dx\left\{\frac{1}{2}\,\phi_\upsilon
\left(\overleftarrow{\nabla}A_{\upsilon\upsilon'}
{\nabla}+T_{\upsilon{\upsilon}'}\right)\phi_{\upsilon'}
\right.\nonumber\\&&\mbox{}\left.
a^{-d}
\,w(\phi,\psi)
\right\}
\end{eqnarray}
with
\begin{eqnarray}
\left(A_{\upsilon\upsilon'}\right)=\frac{-1}{2d\,a^d}\sum_{\boldsymbol{j}\ne \mathbf{0}}
\left(\begin{array}{cc}
{\mathcal{P}}_{\mathbf{0}\boldsymbol{j}}&{\mathcal{R}}_{\mathbf{0}\boldsymbol{j}}\\
{\mathcal{R}}_{\mathbf{0}\boldsymbol{j}}&{\mathcal{Q}}_{\mathbf{0}\boldsymbol{j}}
\end{array}\right)\boldsymbol{r}_{\mathbf{0}\boldsymbol{j}}^2
\end{eqnarray}
and
\begin{eqnarray}
\left(T_{\upsilon\upsilon'}\right)=a^{-d}\sum_{\boldsymbol{j}\ne \mathbf{0}}
\left(\begin{array}{cc}
{\mathcal{P}}_{\mathbf{0}\boldsymbol{j}}&{\mathcal{R}}_{\mathbf{0}\boldsymbol{j}}\\
{\mathcal{R}}_{\mathbf{0}\boldsymbol{j}}&{\mathcal{Q}}_{\mathbf{0}\boldsymbol{j}}
\end{array}\right)\,.
\end{eqnarray}
Here $\overleftarrow{\nabla}$ acts to the left, while $\nabla$
acts as usual to the right. Derivatives of higher than second order
have been dropped.

For the BEG model (\ref{HBEG}) with NN interactions and positive values of the
NN bonds $J$, $K$, and $L$, the 
matrix $(A_{\upsilon\upsilon'})$ is positive definite provided $J+K>2L$.
To see this note that
\begin{eqnarray}
(A_{\upsilon\upsilon'})=\frac{1}{a^d\,d}\,\frac{d}{dq^2}\,
\tilde{\boldsymbol{M}}(\boldsymbol{q})\big|_{\boldsymbol{q}=0}\,
\end{eqnarray}
where
\begin{equation}
\tilde{\boldsymbol{M}}(\boldsymbol{q})\equiv\left(\sum_{\boldsymbol{j}\ne\mathbf{0}}
M_{\mathbf{0}\boldsymbol{j}}^{\upsilon\upsilon'}\,
\mathrm{e}^{i\boldsymbol{q}\cdot\boldsymbol{r}_{\mathbf{0}\boldsymbol{j}}}\right)
\end{equation}
is the Fourier transform of $\boldsymbol{M}$. But
\begin{equation}\label{mder}
\frac{d}{dq^2}\,
\tilde{\boldsymbol{M}}(\boldsymbol{q})=-\tilde{\boldsymbol{M}}(\boldsymbol{q})\cdot
\frac{d\tilde{\boldsymbol{M}}(\boldsymbol{q})^{-1}}{dq^2}
\cdot\tilde{\boldsymbol{M}}(\boldsymbol{q})
\end{equation}
and
\begin{equation}\label{minvder}
-\frac{d}{dq^2}\tilde{\boldsymbol{M}}(\boldsymbol{q})^{-1}\big|_{\boldsymbol{q}=\boldsymbol{0}}=
a^2\left(\begin{array}[T]{cc}J&L\\L&K\end{array}\right).
\end{equation}
The latter is positive definite because of our assumption $J+K>2L$.
In conjunction with (\ref{mder}) it follows that
$d\tilde{\boldsymbol{M}}(\boldsymbol{q})/dq^2$ and hence $(A_{\upsilon\upsilon'})$
are also positive definite. Thus $(A_{\upsilon\upsilon'})$
provides a metric, and the quadratic form
\begin{equation}
\left(T_{\upsilon\upsilon'}+a^{-d}\,
\frac{\partial^2\,w(\phi,\psi)}{\partial\phi_\upsilon\,
\partial\phi_{\upsilon'}}
\Big|_{\phi=\psi=0}
\right)\,\phi_\upsilon(\boldsymbol{x})\,\phi_{\upsilon'}(\boldsymbol{x})
\end{equation}
can be diagonalized by a similarity transformation
\begin{equation}
\varphi_\upsilon(\boldsymbol{x}) =U_{\upsilon\upsilon'}\,\phi_{\upsilon'}(\boldsymbol{x})\;,
\end{equation}
where $(U_{\upsilon\upsilon'})\equiv\boldsymbol{U}$ is orthogonal with
respect to the metric $(A_{\upsilon\upsilon'})$:
\begin{equation}
\boldsymbol{U}^\mathrm{T}\cdot\left(A_{\upsilon\upsilon'}\right)
\cdot\boldsymbol{U}=\boldsymbol{1}\;.
\end{equation}

For vanishing $L$, $(A_{\upsilon\upsilon'})$ is diagonal, but for $L\ne 0$ it is
not and the quadratic part of the Hamiltonian (\ref{cft1}) must be diagonalized
by such a linear transformation. Thus even at this Gaussian level of the theory,
the fields $\phi$ and $\psi$ `mix', i.~e., the order parameter becomes
a linear combination of $\phi$ and $\psi$, a feature that is expected quite generally
in the nonsymmetric case $L\ne 0$, $H\ne 0$. 

To compute the interaction constants $A_{\upsilon\upsilon'}$,
 we expand in $L$, keeping only contributions
up to first order in $L$. This yields 
\begin{equation}
A=\frac{1}{2d\,a^d\,M_0}\,\sum_{\boldsymbol{j}\ne\mathbf{0}}
\left[\boldsymbol{J}\,(M_0+\boldsymbol{J})^{-1}
\right]_{\mathbf{0}\boldsymbol{j}}\,\boldsymbol{r}_{\mathbf{0}\boldsymbol{j}}^2
+O{\left(L^2\right)}\;,
\end{equation}
\begin{equation}
B=\frac{1}{2d\,a^d\,M_0}\,\sum_{\boldsymbol{j}\ne\mathbf{0}}
\left[\boldsymbol{K}\,(M_0+\boldsymbol{K})^{-1}
\right]_{\mathbf{0}\boldsymbol{j}}\,\boldsymbol{r}_{\mathbf{0}\boldsymbol{j}}^2
+O{\left(L^2\right)}\;,
\end{equation}
and
\begin{eqnarray}
e_{11}&=&-\frac{1+O{\left(L^2\right)}}{2d\,a^d\,M_0}\,
\sum_{\boldsymbol{j}\ne\mathbf{0}}
\left[(M_0+\boldsymbol{J})^{-1}
\boldsymbol{L}\,(M_0+\boldsymbol{K})^{-1}\right.\nonumber\\
&&\left.\quad\mbox{}+
(M_0+\boldsymbol{K})^{-1}\boldsymbol{L}\,(M_0+\boldsymbol{J})^{-1}
\right]_{\mathbf{0}\boldsymbol{j}}\,\boldsymbol{r}_{\mathbf{0}\boldsymbol{j}}^2\;.
\end{eqnarray}

Finally, we expand  the function $w(\phi,\psi)$ in Eq.~(\ref{cft1}) into powers of
$\phi$ and $\psi$, eliminate the resulting $\phi^3$ and $\psi^3$ terms through the
shifts (\ref{phishift}) and (\ref{psishift}), and drop terms of sufficiently high order.
The resulting continuum Hamiltonian then takes the form specified by Eqs.~(\ref{ham})
and (\ref{H1})--(\ref{H12}), except that the coupling constants $e_{21}$ and $f_{21}$
vanish in the used approximation. However, such interaction terms
will be generated when one coarse-grains to a larger scale by integrating out
the corresponding short wave-length degrees of freedom.

}}


\begin{thebibliography}{10}
\bibitem{rem1}
For background, see \cite{Gri73,FB91,CL95}. A typical phase diagram with such a
  critical end point is shown in Fig.\ 1 of \cite{FB91}, for example.
\bibitem{rem:Ab2}
In a RG analysis of the asymptotic behavior at the critical line or end point,
  it would be most convenient to set $A=1$ and to transform $b_2$ to the value
  $\pm1$ (depending on whether the original $b_2$ is positive or negative). The
  main reason we do not do this here is that we wish to retain $b_2$ as a
  variable for use in our analysis of the bulk phase diagram.
\bibitem{rem:redproof}
See Sec.\ III.G.2 of \cite{Weg76}. Wegner's elegant proof utilizes some general
  sophisticated RG formalism. For the proof of statement (\ref{redop}) this is
  not necessary; one can proceed in a more elementary fashion, by considering
  the action of $\mathfrak{G}_\mathrm{tra}[\Upsilon]$ on
  $\mathcal{H}_\mathrm{G}[\psi]$ for arbitrary monomials $\Upsilon$ in $\psi$
  and by writing $\mathcal{O}$ as a sum of integrals ${\int_\Omega}$ of such monomials.
\end{thebibliography}
\end{document}